\documentclass[universe,article,accept,moreauthors,pdftex]{Definitions/mdpi} 
\newcolumntype{C}[1]{>{\PreserveBackslash\centering}m{#1}}
\newcolumntype{R}[1]{>{\PreserveBackslash\raggedleft}m{#1}}
\newcolumntype{L}[1]{>{\PreserveBackslash\raggedright}m{#1}}
\graphicspath{{./Definitions/}}
\usepackage{relsize}
\usepackage{blindtext}
\usepackage{enumitem}
\usepackage{graphicx}
\usepackage{bm}
\usepackage{cancel}
\usepackage{epsfig}
\usepackage{dcolumn}
\usepackage{amsmath}
\usepackage{amssymb}
\usepackage{hhline}
\usepackage{adjustbox}
\usepackage{enumerate}
\usepackage[utf8]{inputenc}
\usepackage{caption}
\usepackage{longtable}
\usepackage{rotating}
\usepackage{tablefootnote}
\usepackage{float}

\usepackage{nameref}

\def\aaps{{Astron.\@  \@Astroph.\ }}
\def\apss{{Astroph.\@ \& \@Space \@Science\ }}

\def \beq  {\begin{equation}}
\def \eeq  {\end{equation}}
\def \ber  {\begin{eqnarray}}
\def \eer  {\end{eqnarray}}

\firstpage{1} 
\pubvolume{9}
\issuenum{7}
\articlenumber{317}
\pubyear{2023}
\copyrightyear{2023}
\externaleditor{Academic Editor: Lorenzo Iorio}
\datereceived{10 June 2023} 
\daterevised{24 June 2023} % Comment out if no revised date
\dateaccepted{27 June 2023} 
\datepublished{30 June 2023} 
\hreflink{https://doi.org/10.3390/\linebreak universe9070317} % If needed use \linebreak
\Title{Effects of a Late Gravitational Transition on Gravitational Waves and Anticipated Constraints} %Please confirm title change.
\TitleCitation{Effects of a Late Gravitational Transition on Gravitational Waves and Anticipated Constraints}

\Author{ {Evangelos Achilleas Paraskevas\orcidA} %MDPI: \hl{} %Please carefully check the accuracy of names and affiliations. Changes will not be possible after proofreading. the author name is the same as in redmine but not the same as in submission system, please confirm % The author's complete name is Evangelos Achilleas Paraskevas, as stated in this context.
and {Leandros Perivolaropoulos\orcidB} *}

\AuthorNames{Evangelos Achilleas Paraskevas, Leandros Perivolaropoulos}
\AuthorCitation{Paraskevas, E.A.; Perivolaropoulos, L.}
\address[1]{Department of Physics, University of Ioannina, 45110 Ioannina, Greece; e.paraskevas@uoi.gr}

\corres{\hangafter=1 \hangindent=1.05em \hspace{-0.82em}Correspondence: leandros@uoi.gr}
\newcommand{\newc}{\newcommand}

\newcommand{\be}{\begin{equation}}
\newcommand{\ee}{\end{equation}}
\newcommand{\ba}{\begin{eqnarray}}
\newcommand{\ea}{\end{eqnarray}}
\newcommand{\bea}{\begin{eqnarray*}}
\newcommand{\eea}{\end{eqnarray*}}
\newc{\D}{\partial}
\newc{\ie}{{i.e.,} }
\newc{\eg}{{e.g.,} }
\newc{\etc}{{etc.} }
\newc{\etal}{{et al.}}
\newc{\lcdm}{$\Lambda$CDM }
\newc{\lcdmnospace}{$\Lambda$CDM}
\newc{\wcdm}{$w$CDM }
\newc{\plcdm}{Planck18/$\Lambda$CDM }
\newc{\plcdmnospace}{Planck18/$\Lambda$CDM}
\newc{\omom}{$\Omega_{0m}$ }
\newc{\omomnospace}{$\Omega_{0m}$}

\newc{\ra}{\Rightarrow}
\newc{\baodv}{$\frac{D_V}{r_s}$ }
\newc{\baodvnospace}{$\frac{D_V}{r_s}$}
\newc{\baoda}{$\frac{D_A}{r_s}$ } 
\newc{\baodanospace}{$\frac{D_A}{r_s}$}
\newc{\baodh}{$\frac{D_H}{r_s}$ }
\newc{\baodhnospace}{$\frac{D_H}{r_s}$}

\abstract{ We investigate the evolution of gravitational waves through discontinuous evolution (transition) of the Hubble expansion rate $H(z)$ at a sudden cosmological singularity, which may be due to a transition of the value of the gravitational constant. We find the evolution of the scale factor and the gravitational wave waveform through the singularity by imposing the proper boundary conditions. We also use existing cosmological data and mock data of future gravitational wave experiments (the ET) to impose current and anticipated constraints on the magnitude of such a transition. We show that mock data of the Einstein Telescope can reduce the uncertainties by up to a factor of three depending on the cosmological parameter considered.}

\begin{document}

\section{Introduction}\label{sec:I}

The standard cosmological model ($\Lambda$CDM) \cite{Planck:2018vyg,Planck:2018lkk,Planck:2018nkj,Planck:2019evm,Planck:2019kim,Planck:2019nip,Planck:2020qil} provides a good fit to
a wide range of cosmological data despite its relatively small number of parameters.
However, some important parameters of the model like the Hubble parameter $H_0$ and parameters related to the growth rate of cosmological perturbations appear to have conflicting best-fit values depending on the type of cosmological observations used for their estimation (for a \mbox{review, see \cite{Perivolaropoulos:2021jda}}). Thus, there are {\it tensions} among the best-fit values of \lcdm cosmological parameters. The most important such tension is the {\it Hubble tension} at a level of $5\sigma$ \cite{DiValentino:2020vnx}. In particular, $Planck$ 2018 CMB measurements and BAO measurements (if combined) give  a best-fit value for the Hubble parameter $H_{0}=(67.4\pm 0.5)\frac{km}{s\, {Mpc} 
}$ \cite{Planck:2018vyg}, while ``local'' measurements using mainly type Ia supernovae as standard candles \cite{Riess:2021jrx} give a best-fit value for the Hubble flow $H_{0}=(73.04\pm1.04)\frac{km}{s\, Mpc}$ \cite{DiValentino:2020vnx,Perivolaropoulos:2022khd,Perivolaropoulos:2021jda}. A wide range of extensions of \lcdm has been implemented to address this tension. These approaches may be divided into three~categories
\begin{itemize}
\item \textit{Early time models} \cite{article01,Poulin:2018cxd,PhysRevD.101.063523,Niedermann:2019olb,Herold:2022iib} (see also \cite{Abdalla:2022yfr} for a review)  that implement new degrees of  freedom (e.g., Early Dark energy \cite{Poulin:2023lkg,article01,Poulin:2018cxd,PhysRevD.101.063523}, New Early Dark Energy \cite{Niedermann:2019olb,Herold:2022iib}, etc.) 
 to decrease the scale of the sound horizon at recombination. 
 This scale can be used as a standard ruler to measure the Hubble parameter \cite{Bernal:2016gxb,Aylor:2018drw} after calibration from the CMB angular power spectrum peak locations \cite{Planck:2018vyg}. These models, however, have two problems: 1. They require significant fine tuning in order to avoid conflict with cosmological observations after the time of recombination \cite{Vagnozzi:2021gjh,Hill:2021yec}. 2. They should also decrease the horizon scale at the time of equal time matter and radiation density, which can also be calibrated using the matter power spectrum \cite{Brieden:2022heh,Bernal:2021yli}.
 
 \item {\it Late {time models}} %MDPI: is italic necessary?.
 % yes please keep it
\cite{Wang:2017cel,Yang:2019qza,Cai:2021wgv,Li:2019yem,DiValentino:2020naf,DiValentino:2020kha,Benevento:2022cql,Heisenberg:2022lob,Hernandez-Almada:2021aiw} that attempt to deform the Planck 18 best-fit \lcdm form of the Hubble free expansion rate $E(z)\equiv H(z)/H_0$ between the time of recombination and the present time so that the present-time value $H(z=0)$ becomes consistent with the SH0ES best-fit value \cite{Riess:2021jrx}. The problem of this class of models is that the significant level of $E(z)$ deformation needed is not consistent with a wide range of other observations constraining the form of $E(z)$ like BAO and SnIa data \cite{Alestas:2021xes,Benevento:2020fev,Alestas:2021luu,DiValentino}.
 \item {{\it Ultra-late time models}} \cite{Marra:2021fvf,Alestas:2020zol,Alestas:2021luu,Perivolaropoulos:2023iqj,Alestas:2021nmi,Perivolaropoulos:2022vql} that investigate the possible presence of either an unaccounted-for systematic effect and/or a change in the fundamental physics taking place during the last $150Myrs$ %MDPI: please confirm if Myrs is unit or not, if is unit, please add space before, and make it normal format instead of italic, the same in whole paper
 %Yes it is unit 150MegaYears. Please use it as is.
 (redshift $z<0.01$) when the calibration of standard candles like SnIa is performed \cite{Perivolaropoulos:2022khd}. The problems of this class of models includes fine tuning since the change in physical laws should not only occur at very late times but also should be consistent with other local observations \cite{Alestas:2022xxm}.
\end{itemize}

In the context of ultra-late time models \cite{Alestas:2021nmi,Alestas:2022xxm,Perivolaropoulos:2021bds,Perivolaropoulos:2022khd}, there is an abrupt transition of the SnIa absolute luminosity and/or of their calibrators (e.g., Cepheids and TRGB \cite{Perivolaropoulos:2021bds}) that becomes realized at a specific distance $d_c\in [10,60]Mpc$ due to either a  physics transition taking place globally in the Universe during the last $150Myrs$ or locally in our local Universe within distances smaller than $60Mpc$. A smoking-gun prediction of this class of models is a sudden change in the SnIa absolute magnitude at a distance $d_c<60Mpc$. Recent analyses have indicated hints for hidden signals in the SH0ES \cite{Perivolaropoulos:2022khd} and Pantheon+ SnIa data that are consistent with such a transition \cite{Perivolaropoulos:2023iqj}. A physical mechanism for the realization of such a transition includes an explicit symmetry breaking in the context of the symmetron-screening mechanism (the asymmetron \cite{Perivolaropoulos:2022txg}) as well as a first-order phase transition in the context of a scalar--tensor theory \cite{Lee:2006vka}. In the context of such a mechanism it would be expected that the Hubble expansion rate $H(z)$ could also receive a step-like discontinuity due to the anticipated transition. Such a transition, however, would be difficult to observe in the context of SnIa standard-candle probes of $H(z)$ due to the peculiar velocity noise that prohibits the precise measurement of $H(z)$ at redshifts $z<0.01$. However, it is still possible to investigate the anticipated constraints on such a physics transition using forthcoming gravitational-wave data and existing cosmological data. This is the goal of the present~analysis.

We thus impose constraints on a transition in the value of the effective gravitational constant $G_{\text{eff}}$, which would induce a sudden change in the Hubble flow rate. If such a change is advocated through a phenomenological parametrization of the form $H(t)\propto [1+\sigma \Theta(t-t_{s})]$, where $\sigma$ is a dimensionless parameter and $\Theta$ is the Heavyside step function, then it is deemed to exhibit the characteristics of a sudden cosmological \mbox{singularity \cite{Barrow_2004,Barrow:2004he,Fernandez-Jambrina:2006qld,Fernandez-Jambrina:2008pwx,Fernandez-Jambrina:2006tkb,Perivolaropoulos:2016nhp}}. For instance, in \textit{sudden cosmological singularities} or \textit{type II}, at the time $t=t_s$ we have a divergence of pressure $t=t_{s}\implies \big{|}p(t_{s})\big{|}\to \infty$ with a finite density $\rho(t_{s})<\infty$ and scale factor $a(t_{s})<\infty$, while the scale factor is continuous at $t_{s}$ \cite{Barrow_2004,Fernandez-Jambrina:2006qld,Fernandez-Jambrina:2006tkb,amendola_tsujikawa_2010}. The Hubble flow 
 is finite, or $H(t_{s})<\infty$, but the its derivative diverges, %Please confirm that the intended meaning has been retained
 or $\dot{H}(t_{s})\to \infty$. In this case, the divergence of pressure leads to a violation of the dominant energy condition \cite{Catto_n_2005}. A \textit{geodesically complete singularity} (like the sudden cosmological singularity \cite{Fernandez-Jambrina:2006qld,Fernandez-Jambrina:2006tkb,Perivolaropoulos:2016nhp}) involves a divergence of
a derivative of the scale factor while $a \neq 0$ and $ a<\infty$ through the moment that occurs. 

\textls[-25]{In this paradigm, the generalized sudden cosmological singularities can be \mbox{considered \cite{Barrow:2004hk},}} where the order-$r$ derivative of the scale factor diverges, leading to divergence of the corresponding $r-1$ derivative of pressure. Our universe could confront a sudden singularity in the near future that would not be the ``end'' of it \cite{Denkiewicz_2012} and could even mimic in the present day standard dark-energy models \cite{Denkiewicz_2012}. 

\textls[-15]{Extensions of $\Lambda$CDM predict the existence of a wide range of singularities \cite{article002,Bamba:2012cp}. These include the \textit{Big Rip singularities} or \textit{type I} \cite{Caldwell:2003vq,Nojiri_2005} where at a finite cosmic time, the scale factor, Hubble flow, energy density and pressure diverge. For these singularities there is \textit{geodesic incompleteness}, except for some values of the barotropic index $w$ for which there is null geodesic completeness \cite{Fernandez-Jambrina:2006tkb}.   The \textit{type III-IV singularities} involve a finite scale factor at the moment of the singularity or $a(t_{s})<\infty$.  \mbox{\textit{Type III singularities (or finite scale factor singularities)} \cite{Stefancic:2004kb,Nojiri:2004pf}}} endure that $ \rho(t_{s})\to \infty,\,\big{|}p(t_{s})\big{|}\to \infty$ in finite cosmic time while the scale factor is discontinuous at $t_{s}$. In addition, \textit{type IV singularities} \cite{Nojiri_2005} admit that $\rho(t_{s})= 0,\,\big{|}p(t_{s})\big{|}= 0$ and the barotropic index, the first derivative of pressure, the second of Hubble \textls[-15]{flow and the third of scale factor diverge. Other singularities include $w$-singularities \mbox{($p(t_{s})=\rho(t_{s})=0$ while $w(t_{s})\to\infty$)
 \cite{PhysRevD.79.063521}},} the Little Rip 
 (all the bound structures dissociate when $a(t_{s})\to \infty,\rho(t_{s})\to \infty$ while $t_{s}\to\infty$) \cite{Frampton_2011} and the Pseudo-Rip (all the bound structures below a corresponding threshold dissociate when $H(t_{s})=cosnt.$ while $t_{s}\to \infty$) \cite{Frampton_2012}. 

\textls[-25]{A measure of the strength of a singularity is the \mbox{\textit{strong curvature criterion by Krolak} \cite{CLARKE1985127}}, which measures 
 the time integrals of the tidal forces along geodesics \cite{Perivolaropoulos:2016nhp} obtained by \textit{Krolak integral}
\cite{Perivolaropoulos:2016nhp,CLARKE1985127}:
$\int^{\tau}_{0}|R^{i}\,_{ 0j0}(\tau')|d\tau'
$ where $\tau$ is the affine parameter along the geodesic. A causal geodesic hits to a \textit{strong curvature singularity} by \textit{Krolak} \cite{CLARKE1985127} for some value $\tau_{s}$ of the affine parameter iff by taking the limit $\tau\to \tau_{s}$ the integral $\int^{\tau}_{0}|R^{i}\,_{ 0j0}(\tau')|d\tau'$ \mbox{diverges\endnote{There is also a completely analogous characterization with the corresponding Tipler \mbox{integral \cite{Tipler:1977zza}}.} \cite{Fernandez-Jambrina:2006tkb,Fernandez-Jambrina:2006qld,Fernandez-Jambrina:2007pbg,Perivolaropoulos:2016nhp}.} } 

A geodesically complete singularity could be induced by a rapid transition of the effective gravitational constant. Such a transition is motivated by the ultra-late time models for the resolution of
the Hubble tension.  It could also help resolve the growth tension, as it would reduce the growth
of density perturbations without affecting the Planck$/\Lambda$CDM background expansion \cite{Marra:2021fvf,Alestas:2021nmi,Perivolaropoulos:2022txg}. An abrupt shift of
$G_{\text{eff}}$ is weakly constrained and the current bounds allow
an abrupt change in $G_{\text{eff}}$ by up to about 5--10 \% at
some cosmological time in the past between the present
time and the time of nucleosynthesis \cite{Perivolaropoulos:2022txg}. This is demonstrated in Table \ref{table1} \cite{Alestas:2021nmi}, where we show some current constraints on an overall change in $G_{\text{eff}}$. 

\begin{table}[H]
\caption{{Constraints on the evolution of the gravitational constant. Methods with star (*) constrain $G_{\ast}$ (Planck-mass-related)  %MDPI: \hl{} %Please make sure that permission has been obtained and there is no copyright issue.%There are no copyright issues since part of information extracted from Table 1 of the corresponding reference, which is duly cited.
, while the rest constrain} $G_{\text{eff}}$ (see Table 1 in {\cite{Alestas:2021nmi}}).}
%\label{tab:summarytable}
%\textbf{Table 1:} 
\label{table1}
%\begin{table}[H] 
%%\caption{This is a table caption. Tables should be placed in the main text near to the first time they are~cited.\label{tab1}}
%%%\newcolumntype{C}{>{\centering\arraybackslash}X}
%%\begin{tabularx}{\textwidth}{CCCC}
%\toprule
%\textbf{Title 1}	& \textbf{Title 2}	& \textbf{Title 3}\\
%\midrule
%Entry 1		& Data			& Data\\
%Entry 2		& Data			& Data \textsuperscript{1}\\
%\bottomrule
%\end{tabularx}
%\noindent{\footnotesize{\textsuperscript{1} Tables may have a footer.}}
%\end{table}
%
%\label{table1}

%\begin{adjustbox}{width=\linewidth,left}
\setlength{\tabcolsep}{2.5mm}

\begin{adjustwidth}{-\extralength}{0cm}
%\centering %% If there is a figure in wide page, please release command \centering
\begin{tabular}{C{9cm}C{1.5cm}C{4cm}C{2cm}} 
 \toprule
  \textbf{Method}  & \boldmath{$\Big|\frac{\Delta G_{\rm eff}}{G_{\rm eff}}\Big|_{max}$} &  \textbf{Time Scale (Yr) }&  \textbf{References} \\
  \midrule 

Hubble diagram SnIa---$1\sigma$ confidence level& 0.1 &$\sim$$10^8$ &\cite{Gaztanaga:2001fh}\\ 

Gravitational waves&$8$ &$5\times10^{-8}$  &\cite{Vijaykumar_2021}  \\ 
Paleontology & $0.1$&$2\times 10^{-11}$  &\cite{Uzan_2003}  \\ 

Big Bang Nucleosynthesis---$2\sigma$ confidence level *  &$0.05$  &$1.4\times 10^{10}$ &\cite{article20,Alvey:2019ctk}\\ 
Anisotropies in CMB---$2\sigma$ confidence level * &$0.095$  &$1.4\times 10^{10}$ &\cite{Wu:2009zb}\\ \bottomrule
\end{tabular}
\end{adjustwidth}
%\end{adjustbox}\\
\end{table}

Since $H^{2}(z)$$\sim$$G_{\text{eff}}(z)$, a sudden change in $G_{\text{eff}}(z)$ would be connected to a transition of the Hubble expansion rate $H(z)$, thus leaving interesting signatures in the cosmological data.  The goal of the present analysis is to identify these signatures in present and future data and impose a constraint on the amplitude and redshift of such a transition from the anticipated gravitational-wave data from the Einstein Telescope (ET) {\cite{Punturo:2010zz,Sathyaprakash:2012jk,Zhao_2011,Taylor:2012db,DAgostino:2019hvh,Belgacem_2018,Maggiore:2019uih,Branchesi:2023mws,Matos:2022uew}} and from other existing data including the Pantheon+ dataset and the BAO-CMB data. {The imprints of a potential transition in fundamental physics during the evolution of the Universe could be detectable within the gravitational-wave spectrum. The current investigation primarily directs its attention towards the ultra-late stages of the Universe, distinct from situations such as the gravitational waves marked by a potential transition between dimensions during the very early stages, as discussed in \cite{Mureika:2011bv}.}

The structure of this paper is the following: In the next section we discuss some theoretical models that could induce such a gravitational transition in the context of scalar--tensor theories. In addition, we discuss the effects of an evolving $G_{\text{eff}}$ on the propagation of gravitational waves with an emphasis on the particular evolution corresponding to an abrupt transition. In Section \ref{subsec:3.1} we construct mock data based on the anticipated \mbox{ET \cite{Zhao_2011,Taylor:2012db,Belgacem_2018,Punturo:2010zz}}, which 
is a third-generation ground-based gravitational-wave observatory designed to significantly improve the sensitivity and frequency range compared to the current second-generation detectors \cite{KAGRA:2013rdx}, such as LIGO \cite{LIGOScientific:2014pky}, Virgo \cite{VIRGO:2014yos} and KAGRA \cite{Aso:2013eba}. 
We then use these data to derive the anticipated level of constraints to be imposed by these data on the amplitude of a gravitational transition. In Sections \ref{subsec:3.2} and \ref{subsec:3.3} we focus on the existing SnIa, BAO and CMB data to impose constraints on such a gravitational transition. We also discuss the anticipated improvement of these constraints in the context of the ET data in Section \ref{subsec:3.4}. Finally, in Section \ref{sec:IV} we conclude, summarize and discuss possible future extensions of our analysis.

\section{Evolving Gravitational Constant and Gravitational Waves in a Cosmological~Background}\label{sec:II}

\subsection{Effective Gravitational Strength in Scalar--Tensor Theories with a Gravitational Transition}\label{sec:II.I}
A natural theoretical setup for the realization of evolution of the strength of gravity includes scalar--tensor gravitational theories. These modified gravity theories are based on the metric $g_{\mu\nu}$ and a scalar field $\Phi$ directly coupled to the Ricci scalar $R$ \cite{Abdalla:2022yfr,Will:1984qgz}. Thus, the gravitational interaction is mediated by both the metric tensor $g_{\mu\nu}$, the scalar field $\Phi$ and an arbitrary coupling that determines the relative strength of the scalar field. The Lagrangian density is written as \be \mathcal{L}=\frac{1}{2}F(\Phi)R-g^{\mu\nu}\partial_{\mu}\Phi\partial_{\nu}\Phi-U(\Phi)+\mathcal{L}_{matter}
\ee
in the Jordan frame \cite{Boisseau:2000pr}, where $F(\Phi)$ is the dimensionless nonminimal coupling function of the scalar field $\Phi$, $U(\Phi)$ is the corresponding potential and    $\mathcal{L}_{matter}$ depends on the metric and on the matter fields (coupled to the metric). In these theories the strength of gravity is determined by the effective gravitational constant  
\begin{equation}\label{equation1}
\frac{G_{\text{eff}}}{G_{N}}=\frac{1}{  F(\Phi)}\frac{2F(\Phi)+4\big{(}\frac{d F(\Phi)}{d\Phi}\big{)}^{2}}{2F(\Phi)+3\big{(}\frac{d F(\Phi)}{d\Phi}\big{)}^{2}} 
\end{equation}
where $G_N$ is the value of Newton's constant in the context of general relativity. We define $G_{\ast}(\Phi)\equiv \frac{G_{N}}{F(\Phi)}$. This is the inverse of the Planck mass squared,  $M_{\ast}^2=G_{\ast}^{-1}$ \cite{Boisseau:2000pr}, and it appears in the Friedmann equations. We now consider a gravitational-strength transition taking place at a redshift $z_s$ in a flat universe, leading to a sudden change in the strength of~gravity
\begin{equation}\label{equation2}
    \frac{G_{\text{eff}}}{G_{N}}=1+\alpha \Theta[z(\Phi)-z_{s}]
\end{equation}

This assumption leads to a discontinous nonminimal coupling function
\be
F^{-1}(\Phi)=1+\alpha\Theta[z(\Phi)-z_{s}]
\label{fphitrans}
\ee
and thus, to a discontinuous Hubble expansion rate
\be
\frac{H^{2}}{H_{0}^{2}}=[1+\alpha\Theta(z-z_{s})][1-\Omega_{m,0}+\Omega_{m,0}(1+z)^3]
\label{hztrans}
\ee

Using this information about $F(\Phi)$, it is straightforward to reconstruct the redshift dependence of $\Phi(z)$ and $U(z)$ using the reconstruction approach of Refs \cite{Esposito-Farese:2000pbo,Boisseau:2000pr,Perivolaropoulos:2005yv}, leading~to 
\begin{equation}\label{equat2}
U(z)=3(1-\Omega_{m,0})[1+\Delta_{0}(\alpha)\delta(z-z_s)]
\end{equation}

\begin{equation}\label{equat4}
\Phi(z)=\Phi_{0}[1+\Delta_{1}(\alpha) \Theta(z-z_{s})]
\end{equation}
where $\Delta_{0}(\alpha)$ and $\Delta_{1}(\alpha)$ are proper functions of $\alpha$. Thus, in this context, it is a spike feature of the potential that can lead to a sudden change in the value of the scalar field $\Phi$ in the context of a flat potential. A similar effect would be induced in the context of a first-order phase transition involving the decay of a false vacuum of the scalar field potential.

\subsection{Luminosity Distance from Gravitational Waves in Modified Gravity}

In modified gravity theories, the luminosity distance as measured with gravitational waves may deviate from the standard general relativistic case due to several factors (\mbox{see \cite{Nishizawa_2018,Arai:2017hxj}):}

\begin{itemize}
\item \textit{Modifications to the cosmological expansion history:} Some modified gravity theories predict a different cosmic expansion history compared to general relativity, which can lead to changes in the luminosity distance--redshift relationship. As a result, the measured luminosity distance using gravitational waves may differ in these \mbox{theories \cite{Linder_2018}.}

\item \textit{Additional gravitational-wave polarization modes:} In some modified gravity theories, gravitational waves can have additional polarization modes \cite{Eardley:1973zuo}, such as scalar or vector modes, in addition to the standard tensor modes found in general relativity. The presence of these additional modes can affect the amplitude and phase evolution of the gravitational-wave signal, leading to a different measured luminosity distance.

\item \textit{Propagation effects:} The propagation of gravitational waves in modified gravity theories can be affected by changes in the effective gravitational constant or the presence of additional fields. These effects can alter the gravitational-wave amplitude and, consequently, the inferred luminosity distance \cite{Belgacem:2017ihm,Nishizawa_2018}.
\end{itemize}

By comparing the luminosity distance measurements obtained from gravitational waves with those obtained from electromagnetic observations, such as type Ia supernovae or galaxy surveys, it is possible to test and constrain modified gravity theories \cite{10.1093/acprof:oso/9780198570745.001.0001,10.1093/oso/9780198570899.001.0001}. Any significant discrepancy between the two sets of measurements could provide evidence for deviations from general relativity and help identify the correct gravitational theory.

The luminosity distance in general relativity plays a crucial role in determining the relationship between the intrinsic luminosity of an astrophysical source and its observed flux. For both electromagnetic and gravitational-wave signals, the amplitude of the signal is expected to decay inversely with the luminosity distance. In general relativity, the electromagnetic and gravitational-wave luminosity distances are the same, as they both depend in the  same way on the underlying cosmological model \cite{10.1093/acprof:oso/9780198570745.001.0001,10.1093/oso/9780198570899.001.0001}.

However, in some modified gravity theories, the effective gravitational constant $G_{\text{eff}}$ can evolve with time or cosmic redshift $z$. This can lead to a difference in the propagation of gravitational waves compared to electromagnetic waves, and thus, a difference in the luminosity distances \cite{Linder_2018}.

To derive the relation between the gravitational-wave and electromagnetic luminosity distances, we can use the following assumptions:

\begin{itemize}
\item The energy carried by gravitational waves is proportional to the square of the amplitude of the wave, which, in turn, is proportional to the time-varying effective gravitational constant $G_{\text{eff}}(z)$.

\item The energy carried by electromagnetic waves is not affected by the time-varying effective gravitational constant, as it is primarily determined by the electromagnetic interaction.
\end{itemize}

From these assumptions, we can write the ratio of the energy carried by gravitational waves to the energy carried by electromagnetic waves as:

\be
\frac{E_{gw}}{E_{em}} \propto \frac{G_{\text{eff}}(z)}{G_{\text{eff}}(0)}
\ee

Since the amplitude of the signal is inversely proportional to the luminosity distance, the ratio of the energy carried by the waves also relates to the ratio of the square of the luminosity distances:

\be 
\frac{E_{gw}}{E_{em}} \propto \frac{\left(d_{L}^{em}(z)\right)^2}{\left(d_{L}^{\text{gw}}(z)\right)^2}
\ee

Combining these two expressions, we obtain the relation between the gravitational-wave and electromagnetic luminosity distances \cite{Linder_2018,Belgacem:2017ihm}:
\be 
d_{L}^{\text{gw}}(z) = d_{L}^{em}(z) \sqrt{\frac{G_{\text{eff}}(z)}{G_{\text{eff}}(0)}}
\label{gwdl}
\ee

This relation shows that the gravitational-wave luminosity distance depends on the electromagnetic luminosity distance and the ratio of the effective gravitational constant at redshift $z$ to its present-day value. By measuring the luminosity distances of gravitational waves and electromagnetic signals from the same astrophysical sources, we can test and constrain modified gravity theories that predict an evolving gravitational \mbox{constant \cite{Nishizawa_2018,Belgacem_2018,article301,article302,DAgostino:2019hvh,Tasinato:2021wol}}.

\subsection{Gravitational Waves in an Expanding Universe}

In an FRW universe, the metric tensor can be written as the sum of the background metric and perturbations:

\begin{equation}
    g_{\mu\nu}(t, \mathbf{x}) = g_{\mu\nu}^{(0)}(t) + h_{\mu\nu}(t, \mathbf{x})
\end{equation}
where $g_{\mu\nu}^{(0)}(t)$ is the background FRW metric and $h_{\mu\nu}(t, \mathbf{{x} %MDPI: is bold necessary?.% Yes, it is!
})$ represents the perturbations. In a vacuum, the energy--momentum tensor is zero and the linearized Einstein equations for the perturbations reduce to the linearized vacuum equations. To focus on the tensor perturbations, we can decompose the perturbations into scalar, vector and tensor \mbox{modes \cite{dodelson2020modern}.} Tensor perturbations are transverse and traceless, satisfying:

\be 
h_{\mu\nu}k^{\mu} = 0, \quad h^{\mu}_{\mu} = 0
\ee

The tensor perturbations can be expanded in terms of a complete set of polarization tensors $e_{\mu\nu}^{(c)}(\mathbf{k})$ with $c = +, \times$:

\be 
h_{\mu\nu}(t, \mathbf{x}) = \sum_{c = +, \times} \int d^{3}k \, h_{c}(t, \mathbf{k}) e_{\mu\nu}^{(c)}(\mathbf{k}) e^{i\mathbf{k} \cdot \mathbf{x}}
\ee

Substituting the perturbed metric into the linearized Einstein equations and projecting out the tensor modes using the polarization tensors, we obtain a wave equation for each polarization mode \cite{dodelson2020modern}:

\be
\ddot{h}_{c}(t, \mathbf{k}) + 3 \frac{\dot{a}}{a} \dot{h}_{c}(t, \mathbf{k}) + \frac{k^2}{a^2} h_{c}(t, \mathbf{k}) = 0, \quad (c = +, \times)
\label{eq10}
\ee

This wave equation describes the evolution of the tensor perturbations in the FRW vacuum. The first term represents the acceleration of the perturbation, the second term accounts for the effect of the expansion of the universe (Hubble damping) and the third term corresponds to the spatial gradient term (restoring force).

The solutions to this equation can be used to study the generation and evolution of gravitational waves in cosmology, as tensor perturbations represent the propagation of gravitational waves in the FRW background. By analyzing the solutions to \mbox{Equation (\ref{eq10}),} we can gain insights into the behavior of gravitational waves in various cosmological scenarios, such as the early universe, inflation and structure formation \cite{dodelson2020modern}.

In the present subsection we focus on the predicted evolution in the presence of a gravitational transition that induces a transition in the background Hubble expansion rate. We start from a simple power-law evolution of the scale factor.

In the presence of a cosmological fluid with an equation of state $p = w\rho$, where $w$ is a constant parameter, the Friedmann equation can be written as:

\begin{equation}
H^2 = \frac{8 \pi G}{3} \rho
\end{equation}

Using the relation $\rho \sim a^{-3(1+w)}$, we can deduce the scale factor's behavior:

\begin{equation}
a(t) \sim t^{\frac{2}{3(1+w)}}
\end{equation}

The different values of $w$ correspond to various types of cosmological fluids or energy components:

\begin{itemize}
\item $w = 0$: Dust
\item $w = \frac{1}{3}$: Radiation
\item $w = -\frac{1}{3}$: Curvature or cosmic strings
\item $w = -\frac{2}{3}$: Domain walls
\item $w \to -1$: Vacuum energy
\end{itemize}

In order to solve the differential Equation \eqref{eq10}, we assume the power-law form of the scale factor $a(t)=\big{(}\frac{t}{t_{s}}\big{)}^{\beta}$ and we have:

\begin{equation}\label{relation4}
\ddot{h}_{c}(t,k) + 3 \frac{\beta}{t}\dot{h}_{c}(t,k) + \frac{k^{2}t_{s}^{2\beta}}{t^{2\beta}}h_{c}(t,k) = 0
\end{equation}

The solution may be expressed in terms of the spherical Bessel functions as:

\begin{equation}\label{relation3}
h_{c}(t,k) = \frac{t^{1-\beta}}{a(t)}\left[A_{k}\, j_{-\frac{1}{2}+\frac{1-3\beta}{2(\beta-1)}}\left(\frac{kt_{s}^{\beta}}{1-\beta} t^{1-\beta}\right) + B_{k}\, y_{-\frac{1}{2}+\frac{1-3\beta}{2(\beta-1)}}\left(\frac{kt_{s}^{\beta}}{1-\beta} t^{1-\beta}\right)\right]
\end{equation}
where $j_{n},y_{n}$ are written in terms of the Bessel functions of the \textit{first} and \textit{second} kind $\mathcal{J}_{n},\,\mathcal{Y}_{n}$, respectively, via the relations $j_{n}(x)=\sqrt{\frac{\pi}{2x}}\mathcal{J}_{n+\frac{1}{2}}(x)=\frac{x^n}{(2n+1)!!}\Sigma_{s=0}^{\infty}\frac{(-1)^s}{s!(n+\frac{3}{2})_{s}}(\frac{x}{2})^{2s}$ and also $y_{n}(x)=\sqrt{\frac{\pi}{2x}}\mathcal{Y}_{n+\frac{1}{2}}(x)=-\frac{(2n-1)!!}{x^{n+1}}\Sigma_{s=0}^{\infty}\frac{(-1)^s}{s!(\frac{1}{2}-n)_{s}}x^{2s}$ and $(...)_{s}$ denotes  Pochhammer's symbol \cite{arfken2012mathematical}.

\subsection{The Imprints of a Sudden Cosmological Singularity}

By making the ansatz of the gravitational transition and also that $a(t)=\big{(}\frac{t}{t_{s}}\big{)}^{\beta}$,  the discontinuity at the Hubble flow is written as:

\begin{equation}
     H(t)=\frac{\dot{a}}{a}=[1+\sigma\Theta(t-t_{s})]\frac{\beta}{t}
\end{equation}
where $\sigma$ is a constant that ``measures'' the transition leap. If we solve the differential equation in the distributional sense, we obtain for the scale factor:

\begin{equation}
     a(t)=\bigg{(}\frac{t}{t_{s}}\bigg{)}^{\beta [1+\sigma \Theta(t-t_{s})]}
 \end{equation}

Thus, we define
\begin{equation}
\Tilde{\beta}(t) =\beta [1+\sigma \Theta(t-t_{s})]
\end{equation}

Setting $h_{c,\textbf{k}}(t,\textbf{x})=h_{c}(t,k)e^{i\textbf{k}\cdot \textbf{x}}$ we obtain the ``generalized'' wave equation \cite{DAgostino:2019hvh}:

\begin{equation}\label{relation}
   \ddot{h}_{c}(t,k)+  3 \frac{\tilde{\beta}(t)}{t}\dot{h}_{c}(t,k)+\frac{k^{2}}{\big{(}\frac{t}{t_{s}}\big{)}^{2\tilde{\beta}(t)}}h_{c}(t,k)=0
\end{equation}

The boundary conditions are obtained by integration in the interval $(t_{s}-\epsilon,t_{s}+\epsilon)$ and lead to continuity for both the gravitational-wave amplitude and its time derivative.
To describe the solution of the differential Equation (\ref{relation}), we consider it as comprising two distinct parts, each of which has the form (\ref{relation3}). The first component corresponds to the time interval $t<t_{s}$, during which $\tilde{\beta}$ equals $\beta$. The second component corresponds to the time interval $t>t_{s}$, during which $\tilde{\beta}$ equals $\beta(1+\sigma)$. Both components constitute the complete solution of the differential Equation (\ref{relation}), if the appropriate boundary conditions

\begin{equation}\label{bound0}
h^{(+)}_{c}(t_{s},k)=h^{(-)}_{c}(t_{s},k)
\end{equation}

\begin{equation}\label{bound1}
\dot{h}^{(+)}_{c}(t_{s},k)=\dot{h}^{(-)}_{c}(t_{s},k)
\end{equation}
are imposed. Then, the imprints of the sudden cosmological singularity on the gravitational-wave waveform for different parameters may be obtained. The differential Equation (\ref{relation}) in the two distinct regions of spacetime is expressed in the following manner: if $t<t_{s}$, the differential Equation (\ref{relation}) matches to
\begin{equation}
    \ddot{h}_{c}(t,k) + 3 \frac{\beta}{t}\dot{h}_{c}(t,k) + \frac{k^{2}t_{s}^{2\beta}}{t^{2\beta}}h_{c}(t,k) = 0
\end{equation}
while if $t>t_{s}$, the Equation (\ref{relation}) matches to a different form
\begin{equation}
\ddot{h}_{c}(t,k) + 3 \frac{\beta(1+\sigma)}{t}\dot{h}_{c}(t,k) + \frac{k^{2}t_{s}^{2\beta(1+\sigma)}}{t^{2\beta(1+\sigma)}}h_{c}(t,k) = 0
\end{equation}

The solution in each region can be seen below:
\begin{equation}\label{sol1}
h_{c}(t,k)=
\begin{cases}
    t^{\frac{1-3\beta}{2}}\big{[}A \,\mathcal{J}_{\frac{1-3\beta}{2(1-\beta)}}\big{(}\frac{kt_{s}}{1-\beta}t^{1-\beta}\big{)}+B\, \mathcal{Y}_{\frac{1-3\beta}{2(1-\beta)}}\big{(}\frac{kt_{s}}{1-\beta}t^{1-\beta}\big{)}\big{]}&\, t<t_{s}\\

    t^{\frac{1-3\beta(1+\sigma)}{2}}\big{[}\Gamma\, \mathcal{J}_{\frac{1-3\beta(1+\sigma)}{2[1-\beta(1+\sigma)]}}\big{(}\frac{kt_{s}t^{1-\beta(1+\sigma)}}{1-\beta(1+\sigma)}\big{)}+\Delta\,  \mathcal{Y}_{\frac{1-3\beta(1+\sigma)}{2[1-\beta(1+\sigma)]}}\big{(}\frac{kt_{s}t^{1-\beta(1+\sigma)}}{1-\beta(1+\sigma)}\big{)}\big{]}&\, t>t_{s}
    \end{cases}
\end{equation}

Assuming a system of initial conditions for $h_{c},\dot{h}_{c}$ and the boundary conditions (\ref{bound0}) and (\ref{bound1}), a system of algebraic equations is generated. The model accounts for the propagation of a gravitational wave, as predicted by general relativity.  

\subsection{Effects of a Gravitational Transition: A $\delta$-Function Impulse}
In the context of a scalar--tensor theory, which induces a gravitational transition, the expansion-rate transition may be attributed to the gravitational transition as described at Section \ref{sec:II.I}, i.e.,
 \begin{equation}
     H(t)=\sqrt{\frac{8\pi G_{\text{eff}}(t)}{3}\rho(t)}=\sqrt{\frac{8\pi G_{N}}{3}\rho(t)}\,\big{|}1+\sigma \Theta(t-t_{s})\big{|}
 \end{equation}
 
The effective Newton's constant transition is expressed as $G_{\text{eff}}(t)= G_{N} [1+\sigma \Theta(t-t_{s})]^2$. By using that $\Theta^2=\Theta$ and setting  $\alpha\equiv \sigma(\sigma+2)$, 
\begin{equation}
    G_{\text{eff}}(t)=[1+\alpha \Theta(t-t_{s})]G_{N}
\end{equation}

 Thus, the differential equation of the modified propagation of gravitational waves may be written as
\cite{Saltas:2014dha,Pettorino:2014bka}: 
\begin{equation}\label{eq10a}
  \ddot{h}_{c}(t,k)+ (3+\alpha_{M}) \frac{\dot{a}}{a}\dot{h}_{c}(t,k)+\frac{k^{2}}{a^{2}}h_{c}(t,k)=0, \quad (c=+,\times)
\end{equation}
where $\alpha_{M}=\frac{d \ln G^{-1}_{*}}{d\ln a}=-H^{-1}\frac{\dot{G}_{*}}{G_{\ast}}$, where $G_{*}=M^{-2}_{P}$ (the effective Planck mass) \cite{Gleyzes:2014rba,Belgacem_2019}. Setting $G_*\simeq G_{\text{eff}}=[1+\alpha \Theta(t-t_{s})]G_{\text{eff}}(0)$ we find

\begin{equation}\label{at}
    \alpha_{M}=H^{-1}\frac{\alpha \delta(t-t_{s})}{1+\alpha \Theta(t-t_{s})}
\end{equation}
leading to

\begin{equation}\label{relation1}
   \ddot{h}_{c}(t,k)+  \bigg{[}3+\bigg{(}\frac{\tilde{\beta}}{t}\bigg{)}^{-1}\frac{\alpha \delta(t-t_{s})}{1+\alpha \Theta(t-t_{s})}\bigg{]} \frac{\tilde{\beta}(t)}{t}\dot{h}_{c}(t,k)+\frac{k^{2}}{\big{(}\frac{t}{t_{s}}\big{)}^{2\tilde{\beta}(t)}}h_{c}(t,k)=0
\end{equation}

In this framework, the differential Equation (\ref{relation1}) is integrated within the interval $(t_{s}-\epsilon,t_{s}+\epsilon)$ to derive the boundary conditions, resulting in the gravitational-wave amplitude $h_{c}$ being continuous at $t_{s}$ as

\begin{equation} \label{boundary1}
h^{(+)}_{c}(t_{s},k)=h^{(-)}_{c}(t_{s},k) 
\end{equation}

but the first time derivative of the tensor perturbations exhibits a discontinuity that is induced by the $\delta$-function. The boundary condition of the $\dot{h}_{c}$ at $t_{s}$ is written as follows

\begin{equation}\label{boundary2}
    \lim_{\epsilon\to 0}[\dot{h}_{c}(t_{s}+\epsilon,k)-\dot{h}_{c}(t_{s}-\epsilon,k)]= -   \frac{\alpha}{1+\frac{\alpha}{2}}\dot{h}_{c}(t_{s},k)
\end{equation}

We have set the zero value of the Heaviside function to be $\Theta(0)=\frac{1}{2}$.  

\subsection{Numerical Solutions}
As a working example, we consider a scenario where the value of the scale factor at the present time $t_0$ is set to $a(t_{0})=1$, and we hypothesize a sudden cosmological singularity that occurred in the past $(t_{s}<t_{0})$. In order to incorporate those conditions, we normalize the scale factor as

\begin{equation}\label{scale}
    a(t)=a_{s} \bigg{(}\frac{t}{t_{s}}\bigg{)}^{\beta [1+\alpha\Theta(t-t_{s})]}
\end{equation}

After rescaling to a dimensionless time $\tau=kt$, we can express the value of the scale factor at the moment of singularity, denoted by $a_s$, as $a_{s}=\big{(}\frac{\tau_{s}}{\tau_{0}}\big{)}^{\beta(1+\alpha)}$. Under this transformation, the corresponding differential Equation (\ref{relation}) can be written as:

\begin{equation}\label{diffeq1}
\frac{d^2}{d\tau^2}h_{c}(\tau)+ 3 \frac{\tilde{\beta}(\tau)}{\tau}\frac{d}{d\tau}h_{c}(\tau)+\frac{1}{a^{2}_{s}\big{(}\frac{\tau}{\tau_{s}}\big{)}^{2\tilde{\beta}(\tau)}}h_{c}(\tau)=0
\end{equation}

The appropriate boundary conditions must be applied to obtain the solutions for each scenario. In the case of general relativity, the boundary conditions (\ref{bound0}) and (\ref{bound1}) are necessary, while in the case of scalar--tensor gravity, the boundary conditions (\ref{boundary1}) and (\ref{boundary2}) are required.  

%It is crucial to consider the additional term $\alpha_{M}$ while applying these boundary conditions (\ref{boundary1}) and (\ref{boundary2}). 
The waveform of the system exhibits a sudden change in behavior during the transition from the initial phase, which is represented by the gray waveform and occurs when the scale factor $a$ is less than $a_s$, to  the final phase that occurs when $a$ is greater than $a_s$. At the value of the scale factor $a_s$, a sudden cosmological singularity arises, leading to an abrupt shift in the system's behavior. 

Even though the factor $\frac{\tilde{\beta}}{\tau}$ of the term $\dot{h}$ in differential Equation (\ref{relation}) may be negligible at late times, any sudden change in $\beta$ would result in observable effects at all \mbox{anticipated scales.}

\begin{figure}[H] 

\begin{adjustwidth}{-\extralength}{0cm}
%\centering %% If there is a figure in wide page, please release command \centering
    \centering
    \includegraphics[width=1.2\textwidth]{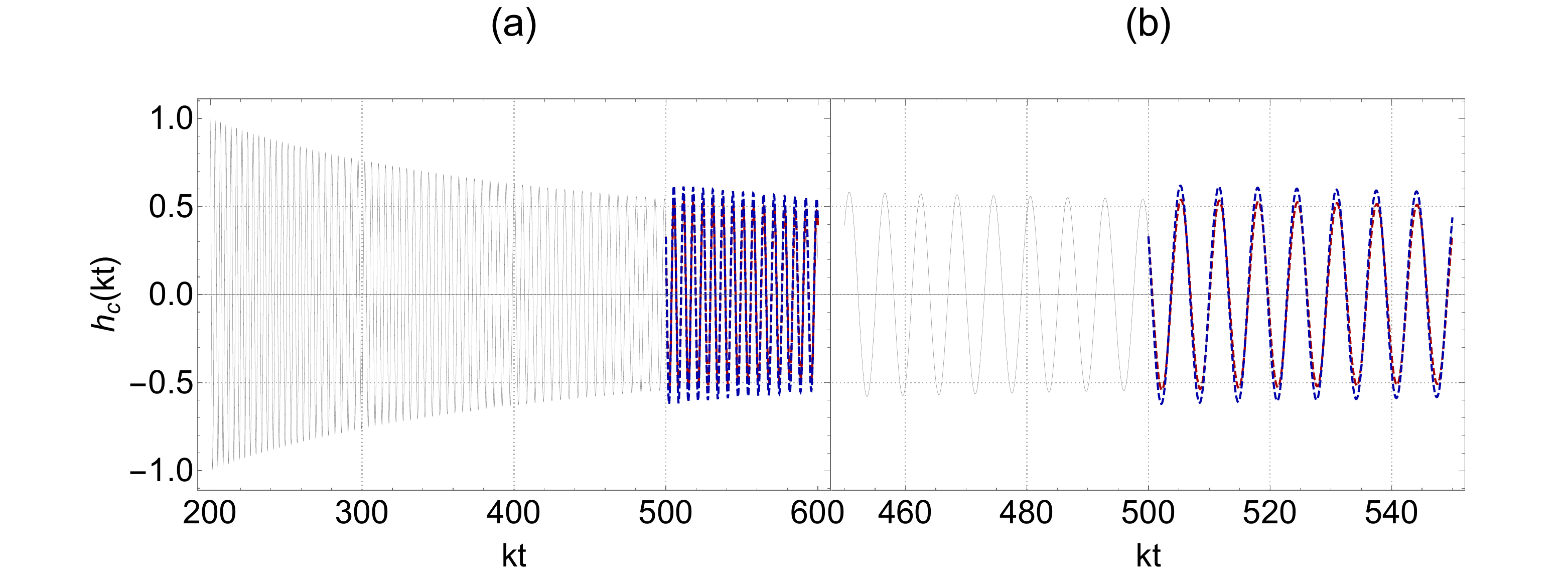}
\end{adjustwidth}
     \caption{ %\hl{There} %MDPI: please add explanation of subfigure.% We added explanations for the subfigures!
  {The} presented figures illustrate a scenario involving a transition from a matter-dominated phase ($\tau<\tau_{s}$) with a fixed value of $\beta=\frac{2}{3}$ to a phase ($\tau>\tau_{s}$) with $\beta=\frac{2}{3}(1+\sigma)$. The dimensionless $\tau$ is plotted along the $x$-axis, and the following parameters for each case,   $\tau_{s}=500,\,\beta=\frac{2}{3},\,\sigma=0.1,\, a_{s}=0.995$, are fixed. We chose  as initial conditions  $h_{c}(\tau=200)=1$ and $h_{c}'(\tau=200)=0$, just for illustration. The \textbf{red}  waveforms correspond to the gravitational waves after the singularity in the context of general relativity ($\alpha_{M}=0$) and the \textbf{blue} waveforms correspond to the propagation of gravitational waves with a modification that is incorporated by the friction term $\alpha_{M}$. The figures, denoted as 1 (\textbf{a}) and (\textbf{b}), depict an identical scenario, albeit corresponding to different ranges of the parameter $\tau\equiv kt$. It is also assumed that $\Theta(0) = \frac{1}{2}$. }
    \label{SFS2}
\end{figure}

\begin{figure}[H] 
 
\begin{adjustwidth}{-\extralength}{0cm}
%\centering %% If there is a figure in wide page, please release command \centering
   \centering
    \includegraphics[width=1.05\textwidth]{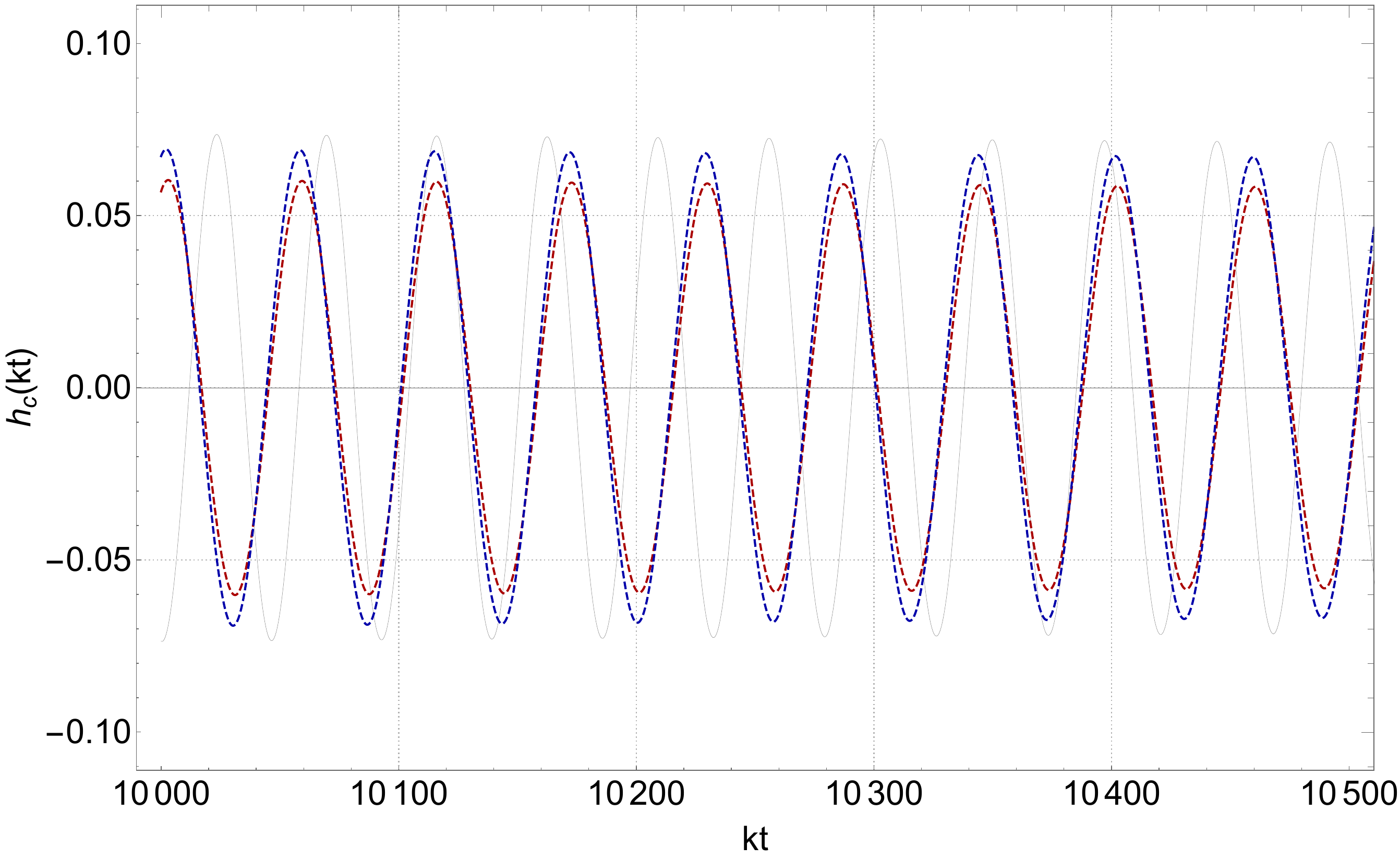}
\end{adjustwidth}
     \caption{ The long-term effects on the amplitude of the gravitational wave due to friction, as opposed to the initial gravitational wave (gray waveform). This behavior is due to a transition from a matter-dominated phase ($\tau<\tau_{s}$) with a fixed value of $\beta=\frac{2}{3}$ to a phase ($\tau>\tau_{s}$) with $\beta=\frac{2}{3}(1+\sigma)$. The dimensionless parameter $\tau$ is plotted along the $x$-axis, and the following parameters are fixed for each case: $\tau_{s}=500,\,\beta=\frac{2}{3},\,\sigma=0.1,\, a_{s}=0.995$. To illustrate this behavior, we selected the initial conditions $h_{c}(\tau=200)=1$ and $h_{c}'(\tau=200)=0$. The \textbf{red} waveforms represent the gravitational waves after the singularity within the context of general relativity ($\alpha_{M}=0$), while the \textbf{blue} waveforms depict the propagation of gravitational waves with a modification, which is incorporated by the extra friction term $\alpha_{M}$ through a gravitational transition. Additionally, we assume that $\Theta(0) = \frac{1}{2}$. }
    \label{SFS3}
\end{figure}

As a concrete example, we use the initial conditions  $h_{c}(\tau=200)=1$ and $h_{c}'(\tau=200)=0$. Moreover, we  assign specific values to the parameters, such as $\beta=\frac{2}{3},\,\tau_{s}=500,\,\sigma=0.1$. We thus find the waveform evolution over the dimensionless rescaled time $\tau$ in  the context of both general relativity and the scalar--tensor theory of gravitational transition frameworks. These are illustrated in Figures \ref{SFS2} and \ref{SFS3}. 

\vspace{-6pt}

The impulse induced by the Hubble flow through the friction term amplifies (or reduces) the amplitude at the moment of the transition (Figure \ref{SFS2}) and results in a slightly faster (or slower) decay of the amplitude compared to the initial phase, depending on the sign of the Hubble rate discontinuity (Figure \ref{SFS3}). In the given context, the value of $\sigma$ is positive, and there is an increase in damping, which arises from the interaction between the gravitational waves and the expanding universe. Therefore, if the damping becomes stronger (weaker), the amplitude of the gravitational wave strain would decay more (less) quickly. The third term in the dynamical equation for gravitational waves affects the wavelength of the oscillating waves. Thus, a sudden increase in the expansion rate would lead to a larger scale factor at later times in the final phase and thus, to a larger gravitational-wave wavelength and period.

In summary, in the general relativistic transition scenario, the cosmological friction term and the third (driving) term in the gravitational wave equation act in different ways, where the former affects the rate of decay of the gravitational wave strain, while the latter affects the wavelength and period of the oscillations.

In the context of modified gravity, which involves modifying the propagation of gravitational waves by introducing an additional term $\alpha_{M}$ \cite{Gleyzes:2014rba,Nishizawa_2018} (illustrated in \mbox{Figures \ref{SFS2} and \ref{SFS3}}), the amplitude of the gravitational waves (blue waveforms) after the gravitational transition through the scalar--tensor theory of gravity is altered compared to the red waveforms, corresponding to transitions in the context of general relativity. This amplitude may increase or decrease depending on the phase of the gravitational wave at the time of the transition and also on the sign of the $\delta$-function impulse determined by the transition amplitude $\alpha$.

\section{Observational Constraints on the Transition Amplitude from Gravitational Waves and Other Cosmological Data}\label{sec:III}

\subsection{Monte Carlo Data: Cosmological Parameters from the Einstein Telescope}\label{subsec:3.1}

The ET \cite{Punturo:2010zz,Sathyaprakash:2012jk,Maggiore:2019uih} is expected to significantly increase the number of detectable standard siren events due to its enhanced sensitivity and broader frequency range compared to current detectors like LIGO \cite{LIGOScientific:2014pky} and Virgo \cite{VIRGO:2014yos}. Although it is difficult to provide exact numbers, as the event rates depend on various factors such as the population of merging compact objects and their distribution in the universe, estimates suggest that the ET could detect thousands of standard sirens per year.

The ET's increased sensitivity will allow it to observe gravitational-wave events from much larger distances, reaching higher redshifts than current detectors \cite{Cai:2017aea,Yang:2023qqz,Fu:2019oll,Cai:2021ooo,Mukherjee:2020mha,LISACosmologyWorkingGroup:2022jok}. While LIGO and Virgo can detect binary neutron star mergers up to a redshift of $z$$\sim$$0.1$, the ET is expected to observe such events up to redshifts of $z$$\sim$$2$ or even higher \cite{LIGOScientific:2016wof,Maggiore:2019uih}. This will enable the ET to probe the expansion history of the universe over a significant fraction of its age, thus deriving constraints on cosmological parameters that are competitive with those obtained with electromagnetic waves.

It is important to note, however, that measuring the Hubble constant and other cosmological parameters using standard sirens not only requires detecting the gravitational wave signal but also identifying an electromagnetic counterpart to obtain the \mbox{redshift \cite{Zhang:2019ylr,Califano:2022syd}}. This may be challenging for high-redshift events, as the associated electromagnetic signals could be fainter and harder to detect. Nevertheless, the significant increase in the number of standard siren events detected with the ET will provide a larger sample for Hubble-constant measurements, even if only a fraction of the events have identifiable electromagnetic counterparts.   A possible method to measure the redshift without the associated electromagnetic signals is to use the statistical redshift estimation, which relies on galaxy catalogues and the probability of finding the source in each galaxy considering its mass and star-formation rate \cite{schutz1986,abbott2017,chen2021}.

The ET will measure cosmological parameters using gravitational waves (GW) from compact binary mergers, such as those involving neutron stars or black holes. One key aspect of this method is the determination of the GW luminosity distance. 

The emitted GW waveform $h_{c}(t)$ resulting from the merger of two compact objects is reliant on various parameters, including the masses and spins of the merging objects, as well as the distance to the source and the inclination angle.  Specifically, in the context of a gravitational wave originating from a compact binary coalescence, the amplitude of the strain in the gravitational wave can be expressed using the post-Newtonian (PN) approximation (the actual relationship can be more complex \cite{Khan:2015jqa,Breschi:2022xnc,Breschi:2022ens,timedomain,Estelles:2021gvs}, especially when considering the full waveform of the gravitational-wave signal), which incorporates PN corrections to the phase $\phi(t)$ but not to the amplitude and can be expressed as \cite{10.1093/acprof:oso/9780198570745.001.0001,10.1093/oso/9780198570899.001.0001, Belgacem_2019}:

\begin{equation}
h_{+}(t)=  \frac{2(1+cos^{2}i)}{d_{L}(z)}  \bigg{(}\frac{G_{N}\mathcal{M}_{c}}{c^{2}}\bigg{)}^{\frac{5}{3}}\bigg{(}\frac{\pi f(t)}{c}\bigg{)}^{\frac{2}{3}}cos\big{[}\phi(t)\big{]}
\end{equation}

\begin{equation}
h_{\times}(t)=  \frac{4cosi}{d_{L}(z)}  \bigg{(}\frac{G_{N}\mathcal{M}_{c}}{c^{2}}\bigg{)}^{\frac{5}{3}}\bigg{(}\frac{\pi f(t)}{c}\bigg{)}^{\frac{2}{3}}\,sin\big{[}\phi(t)\big{]}
\end{equation}

The redshifted chirp mass of a compact binary coalescence system is denoted as $\mathcal{M}_c$ and is related to the chirp mass $M_c$ through the expression $\mathcal{M}_{c}=(1+z)M_{c}$. The chirp mass $M_c$ is a combination of the individual masses, $m_1$ and $m_2$, of the binary system and \mbox{is defined }

\begin{equation}
M_c = \frac{(m_1 \, m_2)^{\frac{3}{5}}}{(m_1 + m_2)^{\frac{1}{5}}}
\end{equation}

The time-varying gravitational-wave frequency $f(t)$ evolves during the inspiral phase of the binary coalescence, and the inclination angle $i$ is defined as the angle between the unit vector normal to the orbital plane and the line of sight \cite{10.1093/acprof:oso/9780198570745.001.0001, Belgacem_2019}. By observing the gravitational waves and comparing the detected signal to theoretical templates, it is possible to extract these parameters, as well as the luminosity distance $d_L$ to the source. In particular, the amplitude of the gravitational waves decreases with the distance to the source. By comparing the calculated intrinsic amplitude (which depends on the chirp mass) with the observed amplitude (which depends on the distance), the luminosity distance to the source can be estimated.

The luminosity distance $d_L$ can be expressed in terms of the scale factor $a(t)$, the Hubble parameter $H(z)$ and the redshift $z$ in a flat universe as:
\be
d_L(z) = (1 + z) \int_{0}^{z} \frac{c}{H(z')}  dz'
\ee
where $c$ is the speed of light and $H(z)$ is the Hubble parameter as a function of the redshift $z$. The Hubble parameter for a flat universe is parameterized by the density parameter $\Omega_{m,0}$ for matter, as
\be
H^2(z) = H_0^2 \left[ \Omega_{m,0} (1+z)^3 + 1-\Omega_{m,0}  \right]
\ee 

Here, $H_0$ is the Hubble constant and $z$ is the redshift. By combining the expressions for $d_L(z)$ and $H(z)$, we can relate the luminosity distance to cosmological parameters:
\be
d_L(z) = (1 + z) \int_{0}^{z} \frac{c}{H_0 \sqrt{\Omega_{m,0} (1+z')^3 + 1-\Omega_{m,0}}}  dz'
\ee

By measuring the luminosity distance $d_L$ from gravitational-wave events and obtaining the redshift $z$ (e.g., from the electromagnetic counterpart), we can fit this expression to a sample of standard siren events to estimate the cosmological parameters, such as the Hubble constant $H_0$, the matter density parameter $\Omega_{m,0}$ and the dark energy density parameter $\Omega_\Lambda$.
\newpage
In order to estimate the anticipated constraints to be imposed on cosmological parameters from luminosity distance measurements of the ET,
we construct a mock dataset of 1000 standard sirens that could be obtained by GW observations of the ET \cite{Belgacem_2018} assuming an underlying \plcdm model. We thus derive Monte Carlo measurements of $d_{L}(z_{i})$ of  binary systems and the corresponding redshifts $z_{i}$ of each binary system. Then, we use these data to fit the Hubble flow $H(z)$ assuming an underlying $\Lambda CDM$ model or a sudden-leap model that has two additional parameters (the transition amplitude and the transition redshift). 

Following \cite{Belgacem_2018}, we generate a catalogue of 1000 events in the redshift range $z \in [0.07,2]$. The interval has a $z_{max}=2$, which leads to an angle-averaged signal-to-noise ratio above \mbox{8 \cite{Belgacem_2018,Zhao_2011}}. There is a minimum redshift cut off, depending on the sensitivity of the detector (ET), which is taken as $z_{min}=0.07$ \cite{Belgacem_2018}, in order to exclude galaxies with peculiar velocities  comparable to their recessional velocities due to the Hubble flow.

 The probability for finding a standard siren with the ET in the redshift range $[z,z+dz]$ is given by the number density function \cite{Belgacem_2018,Zhao_2011}: 
 \begin{equation}
 f(z) dz=N_{0} \frac{4 \pi r_{coal}(z) d_{L}^{2}(z)}{H(z)(1+z)^{3}} dz
 \label{fofz}
 \end{equation}
where $r_{coal}(z)$ is the coalescence rate at redshift $z$ \cite{Zhao_2011}:
 \begin{equation} 
 r_{coal}(z) =
  \begin{cases}
    1+2z      & \quad if \quad z\leq 1 \\
    \frac{15-3z}{4}  & \quad if \quad 1<z<5\\
    0 & \quad if \quad z\geq 5 
  \end{cases}
  \end{equation}

\begin{figure}[H] 
%    \centering
    \includegraphics[width=1\textwidth]{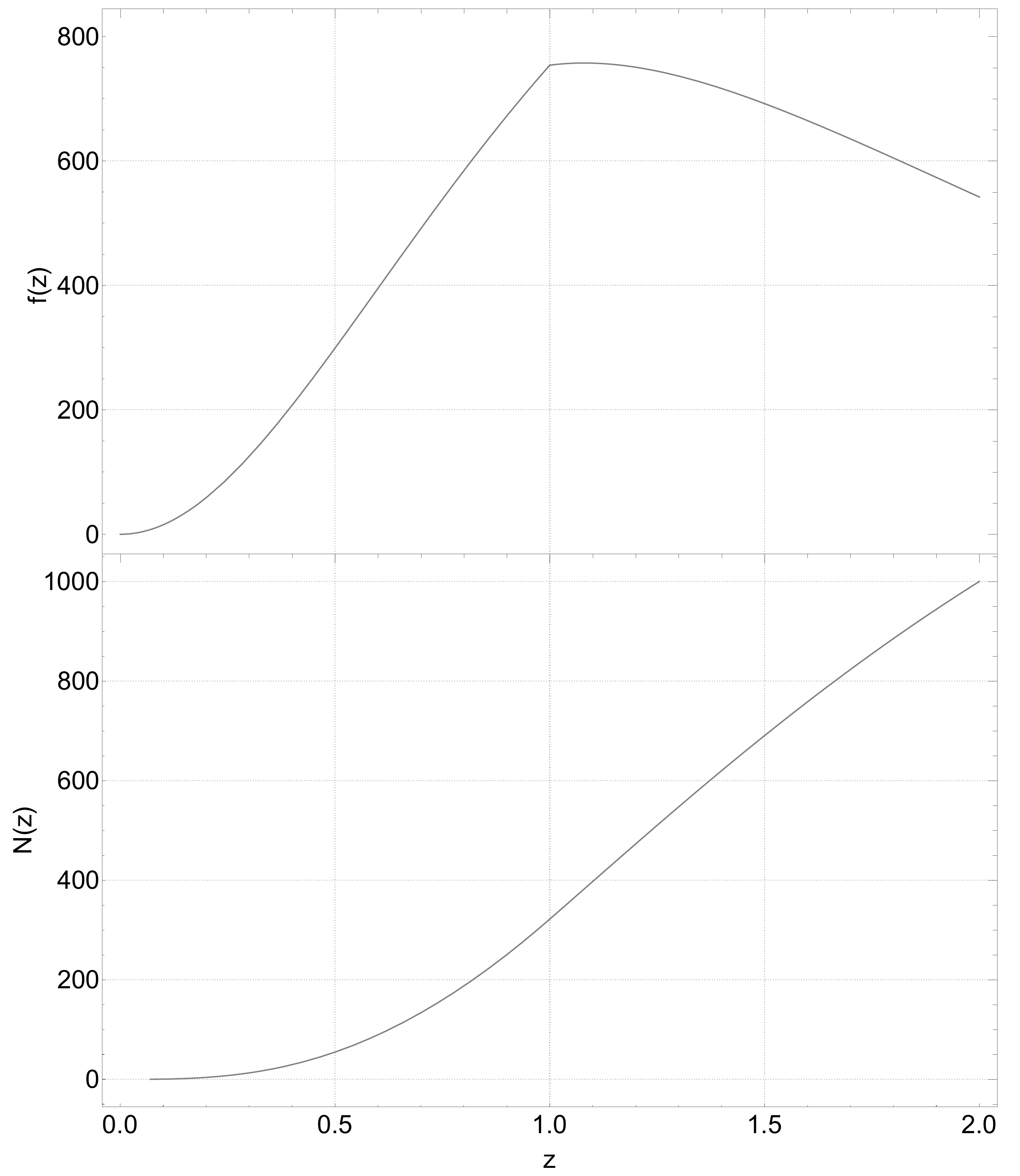}
    \caption{Let $N_0$ be the normalization constant, which is approximately equal to $1.42198 \times 10^{-9}$, and let $f(z)$ be the number density function (see {Figure 8} in \cite{Belgacem_2018} %MDPI: please confirm if is figure citaiton in ref not in this paper. % We confirm that the figure citation in the corresponding reference is included in this paper.
 or {Figure 2 } in \cite{Zhao_2011}%MDPI: please confirm if is figure in ref not in this paper.% We confirm that the figure citation in the corresponding reference is included in this paper.
). We define the non-standard cumulative distribution function $N(z)$. To generate a list of non-uniformly distributed $z_i$ values, we solve the equation $N(z_i) = i$ for $i=1,2,\ldots,1000$, where $i$ is an integer.}
    \label{z1}
\end{figure}

The normalization  constant of Equation (\ref{fofz})  $N_{0}$ is determined by requiring that the integral $N_{0} \int^{2}_{0.07} \frac{4 \pi r(z) d_{L}^{2}(z)}{H(z)(1+z_{i})^{3}} dz=1000$. Using \eqref{fofz}, we thus construct the ``cumulative distribution'' function $N(z)$ (see Figure \ref{z1}) of the 1000 standard sirens through the integral $N(z)=\int^{z}_{0.07}f(z')dz'$ in order to generate the redshifts $z=z_{i}$.  Our approach involves the use of the Newton--Raphson method to solve a set of 1000 equations of the form $N(z_{i})=i$ for $i \in {1, 2, \dots, 1000}$. The objective is to obtain the corresponding redshift $z_{i}$, which lies in the interval $[0.07,2]$, for each equation solved.

In this manner, the redshifts are distributed through the interval $[0.07,2 ]$ with respect to the number density function (\ref{fofz}). In accordance with the discussion presented in \cite{Belgacem_2018}, a set of 1000 sources is generated. The luminosity distance of each source at redshift $z_i$ is obtained by randomly varying normally distributed luminosity distances around a mean value $\bar{d}_{L}(z_{i})$. The value of $\bar{d}_{L}(z_{i})$ is predicted by the hypothesized \plcdm model with $\Omega_{m,0}=0.3166$ and $H_{0}=67.27 \frac{km}{s}Mpc^{-1}$ \cite{Planck:2018vyg}. The resulting luminosity distance is calculated as follows

\begin{equation} 
    \bar{d}_{L}(z_{i})=\frac{c(1+z_{i})}{H_{0}}\int_{0}^{z_{i}}\frac{dz'}{\sqrt{1-\Omega_{m,0}+\Omega_{m,0}(1+z')^{3}}}
\end{equation}

\begin{figure}[H] 
%    \centering
    \includegraphics[width=0.98\textwidth]{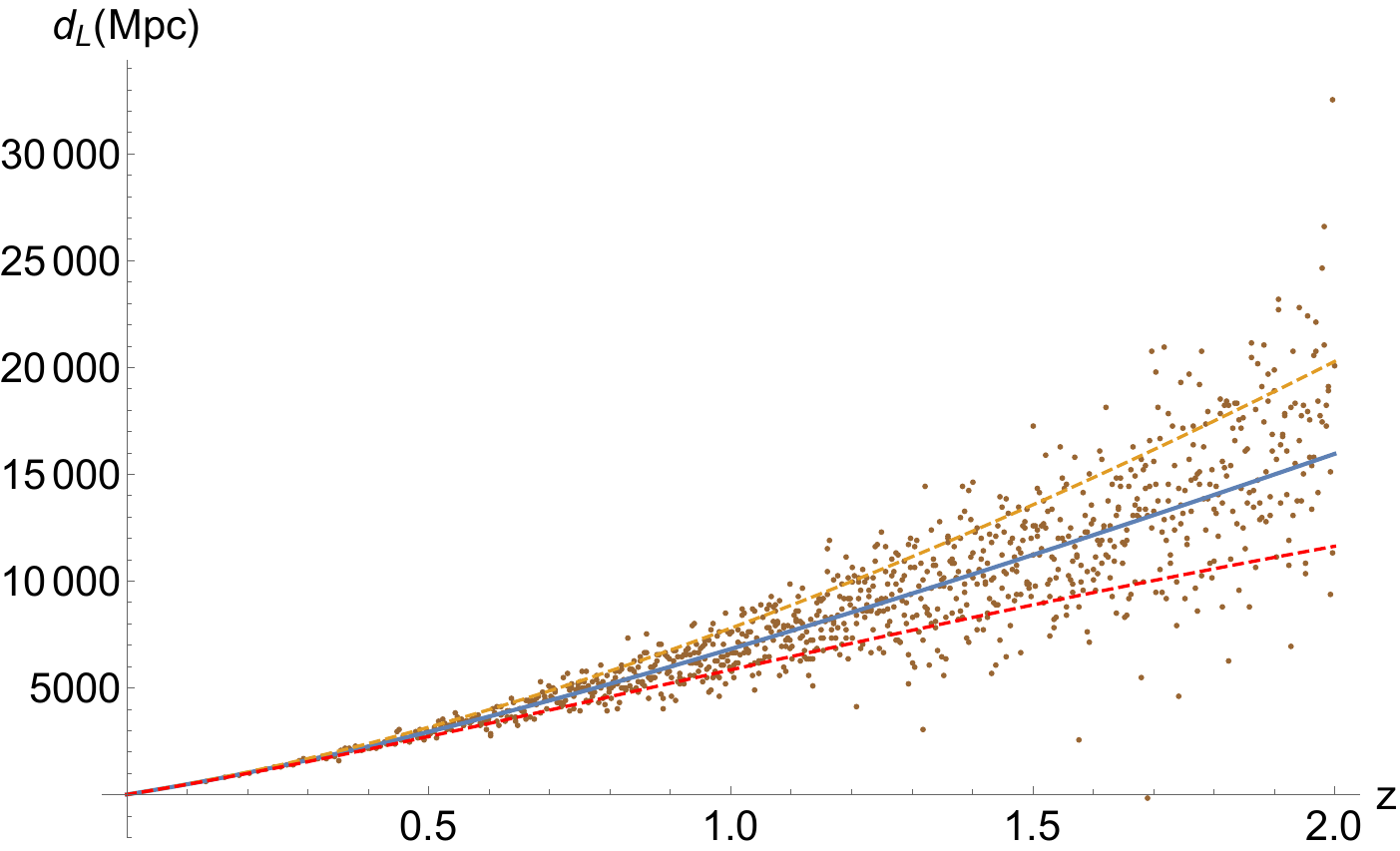}
    \caption{{The} %MDPI: please add comma when number more than five in figure.%  In the figure, there are a maximum of five numbers, so the use of a comma is not necessary.
 cosmology model assumed is $\Lambda$CDM with $\Omega_{m,0}=0.3166$ and $H_{0}=67.27 \frac{km}{s}Mpc^{-1}$, the blue curve corresponds to the mean value or else $\bar{d}_L(z) = (1 + z) \int_{0}^{z} \frac{c}{H_0 \sqrt{\Omega_{m,0} (1+z')^3 + 1-\Omega_{m,0}}}  dz'$, while the yellow and red curves correspond to $\bar{d}_{L}(z)\pm \sigma(z)$,  respectively (see {Figure 7 \cite{Belgacem_2018}} %MDPI: please confirm if is figure citaiton in ref not in maintext, or the figure citaiton order will be wrong order, \hl{} %Please make sure that permission has been obtained and there is no copyright issue.%We have reproduced the referenced figure in our paper, and the corresponding citation is also included in the text.
). The dots constructed are the 1000 standard sirens, which are distributed through the relation (\ref{fofz}).}
    \label{z2}
\end{figure}

We  assume that the luminosity distance $d_{L}(z_{i})$ of each source is normal distributed across an interval of the mean luminosity distance, centered at $\bar{d}_{L}(z_{i})$ with standard deviation \cite{Belgacem_2018,Zhao_2011,Sathyaprakash:2009xt}

  \begin{equation}\label{sigma}
  \sigma_{i}\equiv \Delta d_{L}(z_{i})\approx \bar{d}_{L}(z_{i}) \sqrt{(0.1449 z_{i} - 0.0118 z_{i}^2 + 0.0012 z_{i}^3)^2 + (0.05 z_{i})^2}
  \end{equation} 
where the term $0.05 z_{i}$ is the uncertainty induced by weak lensing \cite{Sathyaprakash:2009xt} and the term $0.1449 z_{i} - 0.0118 z_{i}^2 + 0.0012 z_{i}^3$ is the uncertainty due to instrumental error calculated using Monte Carlo simulations (see \cite{Zhao_2011}). The Gaussian probability density function (PDF), with a standard deviation given by the relation (\ref{sigma}), is written as

\begin{equation}
  PDF(X_{i})=\frac{1}{\sigma_{i}\sqrt{2\pi}} e^{-\frac{1}{2}\big{(}\frac{X_{i}-\bar{d}_{L}(z_{i})}{\sigma_{i}}\big{)}^{2}}
\end{equation}
where $X_{i}$ corresponds to the randomly generated luminosity distance of a standard siren $i$.  Figure \ref{z2} provides a visual representation of a realization of the resulting data \mbox{luminosity distance.}
 
Using these data generated under the assumption of an underlying \plcdm model we can test the type of constraints that would be imposed on a cosmological model involving a discontinuity of $H(z)$ due to a transition of the gravitational constant $G_{\text{eff}}(z)$. The predicted luminosity distance for this class of models is: 

\begin{equation} \label{eq:solve}
    d_{L}(z;\Omega_{m,0},h,\alpha,z_{s})=(1+z)\int_{0}^{z}\frac{c\, dz'}{H_{0}[1+\alpha \Theta(z'-z_{s})]^{\frac{1}{2}}\sqrt{1-\Omega_{m,0}+\Omega_{m,0}(1+z')^{3}}}
\end{equation}
where the Hubble constant is $H_{0}=100h\frac{km}{s\, Mpc}$. In the context of a modified theory of gravity, the hypothesized transition of the gravitational constant would be associated with a mismatch between the luminosity distance $d^{\text{em}}_{L}$ that the light travels %Please confirm that the intended meaning has been retained.
and the gravitational wave luminosity distance $d_{L}^{\text{gw}}$\cite{Belgacem_2018}.

\textls[-15]{In the context of an evolving $G_{\text{eff}}$ the measured $d^{\text{gw}}_{L}(z)$ with standard \mbox{sirens \cite{Belgacem:2017ihm,Belgacem_2018,Belgacem_2019,10.1093/oso/9780198570899.001.0001}}} would lead to a  luminosity distance of the form \cite{Belgacem_2018} shown in Equation (\ref{gwdl}).
% \begin{equation}\label{eq:solve2}
%     d_{L}^{\text{gw}}(z)=d_{L}(z)\sqrt{\frac{G_{\text{eff}}(z)}{G_{\text{eff}}(0)}}
 %\end{equation}

Thus, the luminosity distance measured by gravitational waves may be written as

\begin{adjustwidth}{-\extralength}{0cm}
%\centering %% If there is a figure in wide page, please release command \centering
\begin{equation}\label{eq:solve3}
    d_{L}^{\text{gw}}(z;\Omega_{m,0},h,\alpha,z_{s})=\frac{[1+\alpha \Theta(z-z_{s})]^{\frac{1}{2}}}{(1+z)^{-1}}\int_{0}^{z}\frac{c\,dz'}{H_{0}[1+\alpha \Theta(z'-z_{s})]^{\frac{1}{2}}\sqrt{1-\Omega_{m,0}+\Omega_{m,0}(1+z')^{3}}}
\end{equation}
\end{adjustwidth}
and the cosmological parameters of a cosmological model involving a gravitational transition with amplitude $\alpha$ occuring at redshift $z_s$ may be constrained by minimizing $\chi^{2}$, \mbox{defined as}
 \begin{equation} \label{eq:solve4}
     \chi^{2}_{\text{sirens}}=\sum^{1000}_{i=1}\frac{\big{[}d_{L}^{\text{gw}}(z_{i};\Omega_{m,0},h,\alpha,z_{s})-\bar{d}_{L}(z_{i})\big{]}^{2}}{\sigma_{i}^{2}}\end{equation} 
     
\begin{figure}[H] 
%	    \centering
    \includegraphics[width=0.7\textwidth]{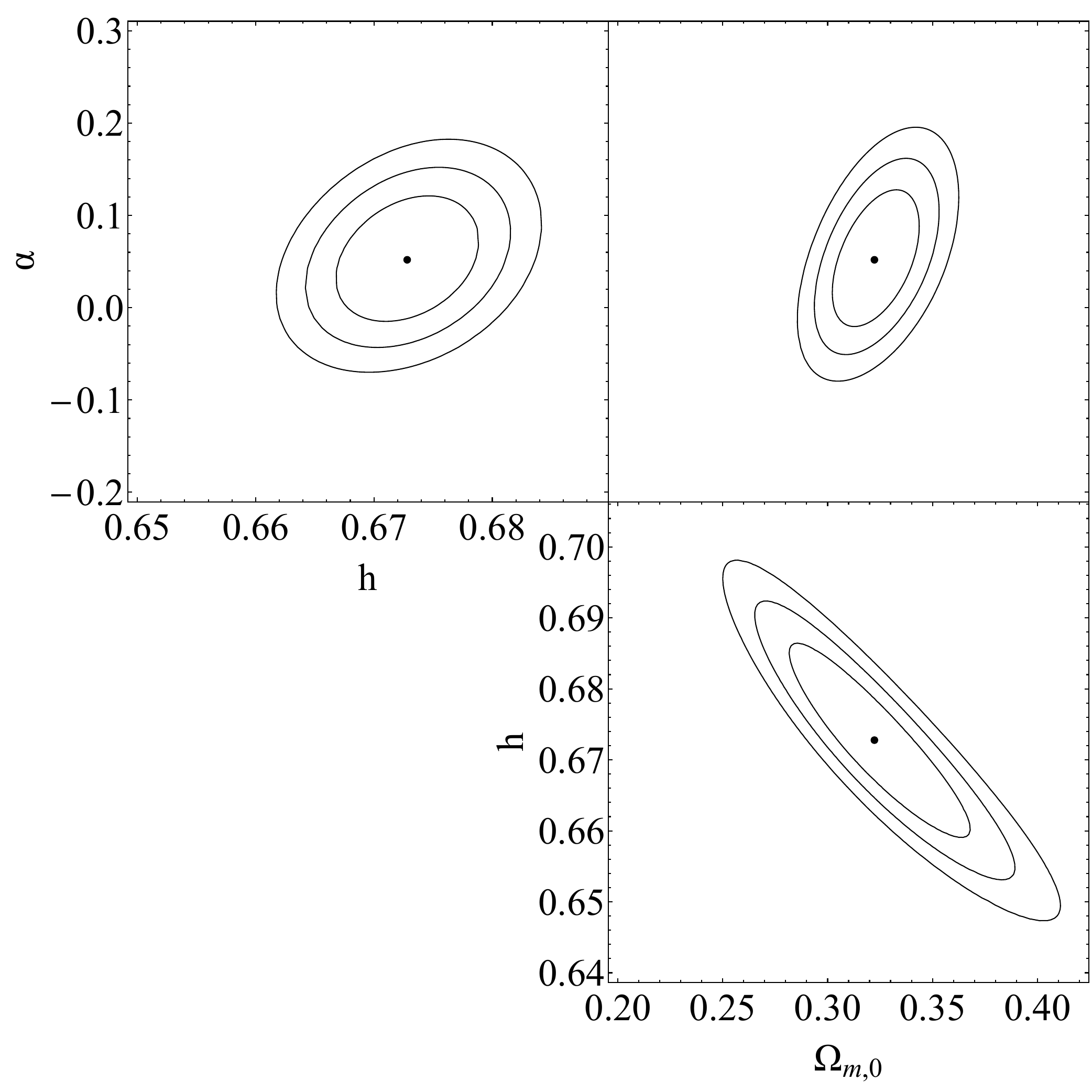}
    \caption{The black contours on the $h-\alpha$, $\Omega_{m,0}-h$ and $\Omega_{m,0}-h$ diagrams represent the projected confidence regions, at $1-3\sigma$, of the standard sirens using mock data. These contours correspond to a transition that occurs at the best-fit value of $a_{s}=0.401 \pm0.397$ and for the corresponding best-fit values each time,  $\Omega_{m,0},h,\alpha$, which can be found in Table \ref{table2}. }
    \label{Data1}
\end{figure}

Since the assumed underlying model is \plcdm and involves no transition, this fit to the Monte Carlo data is  useful only in predicting the level of uncertainties in the anticipated constraints on the sudden-leap model parameters anticipated from the ET data. These predicted constraints are shown in Figure \ref{Data1} and are clearly consistent with $\alpha=0$ since the assumed underlying model is \plcdm\endnote{Some additional contour plots are presented in Appendix \ref{AppendixΑ} (Figure \ref{DataA1}).}. 

\begin{table}[H] 
\scriptsize
\caption{The {$\chi^{2}$ {distributions} %MDPI: plese confirm the table format, we conbine two table. %We confirm!
 and the best-fit parameters accompanied by their corresponding errors can be seen below. The parameters are $M$ (absolute magnitude), $\Omega_{m,0}$ (density parameter for matter), $h$ (parameter related to the Hubble constant $H_{0}$), $\alpha$ (the transition amplitude) and $a_{s}$ (the value of the scale factor at the event of transition).}}
\label{table2}
%\newcolumntype{C}{>{\centering\arraybackslash}X}
\setlength{\tabcolsep}{1.25mm}
\begin{adjustwidth}{-\extralength}{0cm}
%\centering %% If there is a figure in wide page, please release command \centering
\begin{tabular}{C{4cm}C{2cm}C{2cm}C{2cm}C{3cm}C{4cm}}
\toprule
%\textbf{Title 1}	& \textbf{Title 2}	& \textbf{Title 3}\\
   \textbf{Standard Sirens} & \textbf{CMB +BAO} & \textbf{Pantheon+}&  \textbf{Standard Sirens+Pantheon+}&\textbf{CMB+BAO+Pantheon+} & \textbf{Standard Sirens+CMB+BAO+Pantheon+} \\
\midrule
1013.01  & 5.58 &1521.64& 2575.55 & 1570.69 & 2586.44\\ 
%Entry 2		& Data			& Data \textsuperscript{1}\\

\midrule

% 
% \textbf{Table 2:} 

%\begin{adjustbox}{width=\linewidth,left}
%%\begin{tabular}{cccccc}  \toprule
%  \midrule
%
% 
%\hline

%\bottomrule
%\end{tabular}
%\end{adjustbox}
%\begin{adjustbox}{width=\linewidth,left}
%\begin{tabular}{cccccc}  \toprule
%  \midrule
{ \textbf{Data:}} %MDPI: is bold necessary?.%Yes, it is.
 & \boldmath{$M$}  & \boldmath{$\Omega_{m,0}$} & \boldmath{$h$} & \boldmath{$\alpha$}& $\textbf{a}_{s}$ \\
\midrule

\textbf{ Standard Sirens} & --- & 0.323  $\pm$ 0.024& 0.673 $\pm$ 0.008  &0.052 $\pm$  0.123 & 0.401 $\pm$ 0.397\\ 

\textbf{CMB +BAO} & --- & 0.323 $\pm$ 0.009  &0.671 $\pm$ 0.004 &0.008 $\pm$ 0.01 & 0.299 $\pm$ 0.048\\ 

\textbf{Pantheon+} &$-$19.24 $\pm$ 0.03 & 0.325 $\pm$ 0.02 &0.718 $\pm$ 0.023 & 0.057 $\pm$ 0.068 & 0.995 $\pm$ 0.004\\

\textbf{Standard Sirens+Pantheon+} & $-$19.42 $\pm$ 0.01 &0.321 $\pm$ 0.015 &0.678 $\pm$ 0.004 &0.039 $\pm$ 0.025 &0.498 $\pm$ 0.015\\ 

\textbf{CMB +BAO+Pantheon+} & $-$19.43 $\pm$ 0.01 &0.308 $\pm$ 0.009 &0.678 $\pm$ 0.006 &$-$0.004 $\pm$ 0.007 &0.789 $\pm$ 0.2\\

\textbf{Standard Sirens+CMB+BAO+Pantheon+} & $-$19.43$\pm$ 0.01 &0.31 $\pm$ 0.005 &0.677 $\pm$ 0.003 &$-$0.001 $\pm$ 0.005 &0.812 $\pm$ 0.2\\

\bottomrule
\end{tabular}
\end{adjustwidth}
%\noindent{\footnotesize{\textsuperscript{1} Tables may have a footer.}}
\end{table}
\newpage
\subsection{Real Data: BAO+CMB}\label{subsec:3.2}

In the early universe prior to recombination, initial perturbations evolved into overdensities through gravitational interactions with dark matter. Baryonic matter was embedded within these dark matter overdensities and the collapse of these overdensities was followed by radiation-induced overpressure. This overpressure, in turn, generated an expanding sound wave that propagated through plasma at a velocity of 

\begin{equation}
c_{s}=\frac{c}{\sqrt{3(1+R_{s})}}
\end{equation}

Here, $R_{s}\equiv\frac{3 \rho_{b}}{4 \rho_{\gamma}}$, where $\rho_{b}$ represents the baryon density and $\rho_{\gamma}$ represents the photon density. The fluid undergoes damped oscillations in both space and time, wherein the oscillation period depends on the sound speed.  The sound speed $c_{s}$  depends on the density of baryonic matter. When the density of baryons is considerably lower than that of radiation, the sound speed assumes the typical value for a relativistic fluid, i.e., $c_s = c/\sqrt{3}$. However, the introduction of baryonic matter increases the mass of the fluid, leading to a decrease in the sound speed.

If we denote as $r_{s}(z)$ the sound horizon, i.e., the comoving distance traveled by a sound wave from the Big Bang until a corresponding redshift $z$, then \cite{dodelson2020modern,amendola_tsujikawa_2010}
\begin{equation}\label{sound}
r_{s}(z)=\frac{1}{H_{0}}\int_{z}^{\infty} dz' \,\frac{c_{s}(z')}{E(z')}\end{equation} 

The sound horizon $r_{s}(z_{\text{drag}})$  marks the distance over which sound waves propagated. The redshift $z_{\text{drag}}$ denotes the period of the drag epoch, i.e., the epoch when the baryons were released from the Compton drag of photons, which occurred slightly after recombination in the early universe. Eisenstein et al. \cite{Eisenstein:1997ik} obtained a suitable fitting formula for the redshift $z_{\text{drag}}$ as 

\begin{equation}
    z_{\text{drag}}=\frac{1291 (\Omega_{m,0}h^{2})^{0.251}}{1+0.659(\Omega_{m,0}h^{2})^{0.828}}[1+b_{1}(\Omega_{m,0}h^{2})^{b_{2}}]
\end{equation}
where the parameter $b_{1}$ is 
\begin{equation}b_{1}=0.313(\Omega_{m,0}h^{2})^{-0.419} [1+0.607 (\Omega_{m,0}h^{2})^{0.674}]\end{equation}
and  $b_{2}$ is

\begin{equation}b_{2}=0.238 (\Omega_{m,0}h^{2})^{0.223}\end{equation}

The baryon acoustic oscillation (BAO) measurements, which detect the presence of a characteristic scale in the matter distribution, offer a standard ruler that is valuable for deducing the expansion history of the universe and estimating other cosmological parameters. These measurements provide constraints on several quantities, \mbox{specifically \cite{Kazantzidis_2019}:}

\begin{align}
d_{M}\times \frac{r_{s}^{\text{fid}}}{r_{s}},\quad \ d_{V}\times \frac{r_{s}^{\text{fid}}}{r_{s}},\quad \ d_{H}\times \frac{r_{s}^{\text{fid}}}{r_{s}}
\end{align}

In the above equations, $r_{s}^{\text{fid}}$ represents the sound horizon in the fiducial cosmology. The Hubble distance, denoted by $d_{H}$, is a characteristic length scale of the universe and can be expressed as:

\begin{equation}
d_{H}(z)=cH^{-1}(z)
\end{equation}

We define the angular diameter distance, $d_{A}$, and proper motion distance, $d_{M}$. These distances are related as follows:

\begin{equation}\label{dm}
d_{M}(z)=(1+z)d_{A}(z)=\frac{d_{L}(z)}{1+z}
\end{equation}

The related effective distance, $d_{V}(z)$ \cite{SDSS:2005xqv}, is given by the equation:

\begin{equation}
d_{V}(z)=\bigg{[}c z\frac{ d^{2}_{M}(z)}{H(z)}\bigg{]}^{\frac{1}{3}}
\end{equation}

In terms of measurements, there are two possible directions: the line-of-sight dimension and the transverse direction. These measurements involve the ratios $\delta z_{s}\equiv\frac{r_{s}(z_{\star})H(z)}{c}$ and $\theta_{s}(z)\equiv \frac{r_{s}(z_{\star})}{d_{M}(z)}$. The spherically averaged spectrum is connected to the effective distance, which can be expressed as:

\begin{equation}
[\theta_{s}^{2}(z)\delta z_{s}]^{\frac{1}{3}}= z^{\frac{1}{3}}\frac{r_{s}(z_{\star}) }{d_{V}(z)}
\end{equation}

 A fitting formula, developed by Hu and Sugiyama, 1995 \cite{Hu:1995en}, can be utilized to derive the redshift $z_{\star}$ associated with the photon decoupling surface. For values of the baryon density parameter $\Omega_{b}$ in the range $0.0025\leq\Omega_{b}\leq 0.25$ and the matter density parameter $\Omega_{m,0}$ in the range $0.025\leq \Omega_{m,0}\leq 0.64$, $z_{\star}$ can be expressed as:

\begin{equation}
z_{\star}=1048\bigg{[}1+0.00124 (\Omega_{b}h^{2})^{-0.738}\bigg{]}\bigg{[}1+g_{1}(\Omega_{m,0}h^{2})^{g_{2}}\bigg{]}
\end{equation}

\begin{align}
g_{1}=\frac{0.0783(\Omega_{b}h^{2})^{-0.238}}{1+39.5 (\Omega_{b}h^{2})^{0.763}}, \quad \ g_{2}=\frac{0.560}{1+21.1 (\Omega_{b}h^{2})^{1.81}}
\end{align} 

The baryonic acoustic oscillation (BAO) leaves a characteristic imprint on the power spectrum of the cosmic microwave background (CMB) anisotropies that is observed as a series of peaks and troughs. The characteristic angle, $\theta_{A}$, that defines the location of the peaks can be calculated by the following equation \cite{amendola_tsujikawa_2010}:

\begin{equation}
\theta_{A}=\frac{r_{s}(z_{\star})}{d_{M}(z_{\star})}
\end{equation}

 The angular power spectrum of the CMB is decomposed into its multipole moments, where the low multipole moments correspond to the large angular scales and the high multipole moments correspond to the small angular scales. Each multipole $l$ that corresponds to the characteristic angle $\theta_{A}$ can be determined by the following equation \cite{amendola_tsujikawa_2010}:

\begin{equation}
l_{A}=\frac{\pi}{\theta_{A}}=\pi \frac{d_{M}(z_{\star})}{r_{s}(z_{\star})}
\end{equation}

The shift parameter, denoted as R, is a dimensionless parameter that provides a rescaled representation of the ratio between the proper transverse velocity and the observed angular velocity of an object at the photon decoupling surface. It encompasses information related to the comparison of predicted and observed positions of the acoustic peaks in the cosmic microwave background (CMB). Its formal definition is as follows \cite{amendola_tsujikawa_2010}:

\begin{equation}
R\equiv\frac{\sqrt{\Omega_{m,0}H_{0}^{2}}d_{M}(z_{\star})}{c}
\end{equation}

\textls[20]{To impose constraints on the cosmological parameters, we have followed the work} \mbox{of \cite{Theodoropoulos:2021hkk,Zhai:2018vmm,Kazantzidis_2019}}. Throughout the next sections, we incorporate the induced transition in the Hubble flow, which is

\begin{equation}
    H(z;\Omega_{m,0},h,\alpha,z_{s})=H_{0}[1+\alpha \Theta(z-z_{s})]^{\frac{1}{2}}\sqrt{1-\Omega_{m,0}+\Omega_{m,0}(1+z)^{3}}
\end{equation}

\subsubsection{CMB Measurements}

The analysis incorporates Planck data, consisting of temperature and polarization data, along with CMB lensing. Zhai et al. \cite{Zhai:2018vmm} provide the CMB data, which is represented by a data vector and a covariance matrix. The data vector associated with it is expressed \mbox{as follows}

\begin{equation}
v = 
 \begin{pmatrix}
 1.74963\\
301.80845\\
0.02237
 \end{pmatrix}\end{equation}

and also the covariance matrix is written \cite{Zhai:2018vmm}: 
 \begin{equation}
 [C_{CMB}]=10^{-8} \times
 \begin{pmatrix}
  1598.9554 & 17112.007 & -36.311179  \\
  17112.007 & 811208.45 &  -494.79813   \\
  -36.311179  & -494.79813  & 2.1242182   \\
\end{pmatrix}
 \end{equation}

The $\chi^{2}_{\text{CMB}}$ distribution is \cite{Zhai:2018vmm}:

  \begin{equation}
      \chi^{2}_{\text{CMB}}=\textbf{v}^{T}[C_{ CMB}]^{-1}\textbf{v}
      \end{equation}

 where in a flat universe the vector is written as \cite{Zhai:2018vmm} 
\begin{equation}
\textbf{v} = 
 \begin{pmatrix}
  R-1.74963  \\
  l_{A}-301.80845\\
  \Omega_{b}h^{2}-0.02237
 \end{pmatrix}\end{equation}
 
By adopting the luminosity distance (\ref{eq:solve}) and  the proper motion distnce, as defined in (\ref{dm}), then

\begin{equation}
l_{A}=\pi\frac{d_{M}(z_{\star};\Omega_{m,0},h,\alpha,z_{s})}{r_{s}(z_{\star};\Omega_{m,0},h,\alpha,z_{s})}
\end{equation}

and the shift parameter is

\begin{equation}
R\equiv\frac{\sqrt{\Omega_{m,0}H_{0}^{2}}d_{M}(z_{\star};\Omega_{m,0},h,\alpha,z_{s})}{c}
\end{equation}

\subsubsection{BAO Measurements}

The distribution of matter exhibits distinct patterns known as baryonic acoustic oscillations (BAO), which are reflected in the spatial distribution of galaxies. Numerous astronomical surveys, such as the six-degree Field Galaxy Survey (6dFGS), the WiggleZ surveys, the Sloan Digital Sky Survey (SDSS) and the Lyman-alpha (Ly-$\alpha$) survey, have been conducted to map out the distribution of galaxies.

To quantify the BAO measurements, the $\chi^{2}_{\text{BAO}}$ distribution is computed following the methodology outlined in \cite{Theodoropoulos:2021hkk,Escamilla_Rivera_2016}. The expression for $\chi^{2}_{\text{BAO}}$ is given as:

\begin{equation}\label{chibao}
\chi^{2}_{\text{BAO}}=\chi^{2}_{6dFGs,Wiggle}+\chi^{2}_{SDSS}+\chi^{2}_{Ly-a}
\end{equation}

The surveys 6dFGs \cite{article003} and WiggleZ  \cite{Drinkwater:2009sd}  constitute the following distribution \begin{equation}\chi^{2}_{6dFGs,Wiggle}=v_{i}([C_{6dFGs,Wiggle}]^{-1})_{ij}v_{j}\end{equation}

Those surveys give a measurement for the ratio

\begin{equation}
    d_{z}(z)\equiv\frac{r_{s}(z_{\text{drag}})}{d_{V}(z)}
\end{equation}

which constitutes the components 

\begin{equation}v_{i}=d_{z}(z_{i};\Omega_{m,0},h,\alpha,z_{s})\big{|}_{theory}-d_{z}(z_{i})\big{|}_{observed}\end{equation}

where  $d_{z}(z_{i};\Omega_{m,0},h,\alpha,z_{s})$ is defined through  Equations (\ref{eq:solve}) and (\ref{dm}). The corresponding data vectors are \cite{Theodoropoulos:2021hkk}
\begin{equation*}
\begin{pmatrix}
z \\
d_{z} \\
\sigma
\end{pmatrix}=\begin{pmatrix}
0.106 \\
0.336 \\
0.015
\end{pmatrix},\begin{pmatrix}
0.44 \\
0.073 \\
0.031
\end{pmatrix},\begin{pmatrix}
0.6 \\
0.0726 \\
0.0164
\end{pmatrix},\begin{pmatrix}
0.73 \\
0.0592 \\
0.0185
\end{pmatrix}
\end{equation*}

and also the total covariance matrix is \cite{Theodoropoulos:2021hkk}:

 \begin{equation}
 [C_{6dFGs,Wiggle}]^{-1}=
 \begin{pmatrix}
  \frac{1}{0.015^{2}} & 0 & 0 &0 \\
  0 & 1040.3 &  -807.5 & 336.8 \\
  0  & -807.5& 3720.3 & -1551.9   \\
 0  & 336.8& -1551.9 & 2914.9   
 \end{pmatrix}
 \end{equation}

To analyze the SDSS data by constraining  $d_{V}\times \frac{r_{s}^{\text{fid}}(z_{\text{drag}})}{r_{s}(z_{\text{drag}})}$ we utilize the data vectors provided in \cite{Theodoropoulos:2021hkk} and if the value $r_{s}^{\text{fid}}(z_{\text{drag}})=149.28 \,Mpc$, then

\begin{equation*}
\begin{pmatrix}
z \\
\frac{1}{d_{z}} \\
\sigma
\end{pmatrix}=\begin{pmatrix}
0.15 \\
4.465666824 \\
0.1681350461
\end{pmatrix},\begin{pmatrix}
0.32\\
8.4673\\
0.167
\end{pmatrix},\begin{pmatrix}
0.57\\
13.7728\\
0.134
\end{pmatrix}
\end{equation*}
and the corresponding distribution for SDSS data is \cite{Theodoropoulos:2021hkk}\begin{equation}\chi^{2}_{SDSS}=\sum^{3}_{i=1}\frac{\bigg{[}\frac{1}{d_{z}(z_{i};\Omega_{m,0},h,\alpha,z_{s})}\big{|}_{theory}-\frac{d_{V}(z_{i})}{r_{s}(z_{\text{drag}})}\big{|}_{observed}\bigg{]}^2}{\sigma_{i}^{2}}\end{equation}

 The Ly-$\alpha$ survey constrains $  \frac{d_{H}}{r_{s}}$ and $  \frac{d_{M}}{r_{s}}$, respectively (see also \cite{Theodoropoulos:2021hkk}). The corresponding data vectors are \cite{Theodoropoulos:2021hkk} 
\begin{align}    
v_{1}=
\begin{pmatrix}
z\\
\frac{d_{M}}{(1+z) r_{s}(z_{\text{drag}})}\\
\frac{\sigma}{1+z}
\end{pmatrix}
=
\begin{pmatrix}
2.34\\
11.20\\
0.56
\end{pmatrix},\, \quad v_{2}
=
\begin{pmatrix}
z\\
\frac{d_{H}}{r_{s}(z_{\text{drag}})}\\
\sigma\end{pmatrix}=\begin{pmatrix}
2.34\\
8.86\\
0.29
\end{pmatrix}
\end{align}

and they constitute  the covariance matrix 
 
 \begin{equation}\label{cov}
 [C_{Ly-\alpha}]^{-1}=
 \begin{pmatrix}
  \frac{1}{0.56^{2}} & 0  \\
  0 & \frac{1}{0.29^{2}}  \\
 
 \end{pmatrix}
 \end{equation}

  The distribution for Ly-$\alpha$ data $\chi^{2}_{Ly-a}=\textbf{v}^{T}[C_{Ly-\alpha}]^{-1}\textbf{v}$ is constituted by the covariance matrix (\ref{cov}) and the following vector \cite{Theodoropoulos:2021hkk}  
\begin{equation}\textbf{v} = 
 \begin{pmatrix}
\frac{d_{A}(z_{i})}{r_{s}(z_{\text{drag}})}\big{|}_{observed}-\frac{d_{A}(z_{i};\Omega_{m,0},h,\alpha,z_{s})}{r_{s}(z_{\text{drag}};\Omega_{m,0},h,\alpha,z_{s})}\big{|}_{theory} \\
 \frac{d_{H}(z_{i})}{r_{s}(z_{\text{drag}})}\big{|}_{observed}-\frac{d_{H}(z_{i};\Omega_{m,0},h,\alpha,z_{s})}{r_{s}(z_{\text{drag}};\Omega_{m,0},h,\alpha,z_{s})}\big{|}_{theory}
 \end{pmatrix}\end{equation}

\begin{figure}[H] 
%	    \centering
    \includegraphics[width=1\textwidth]{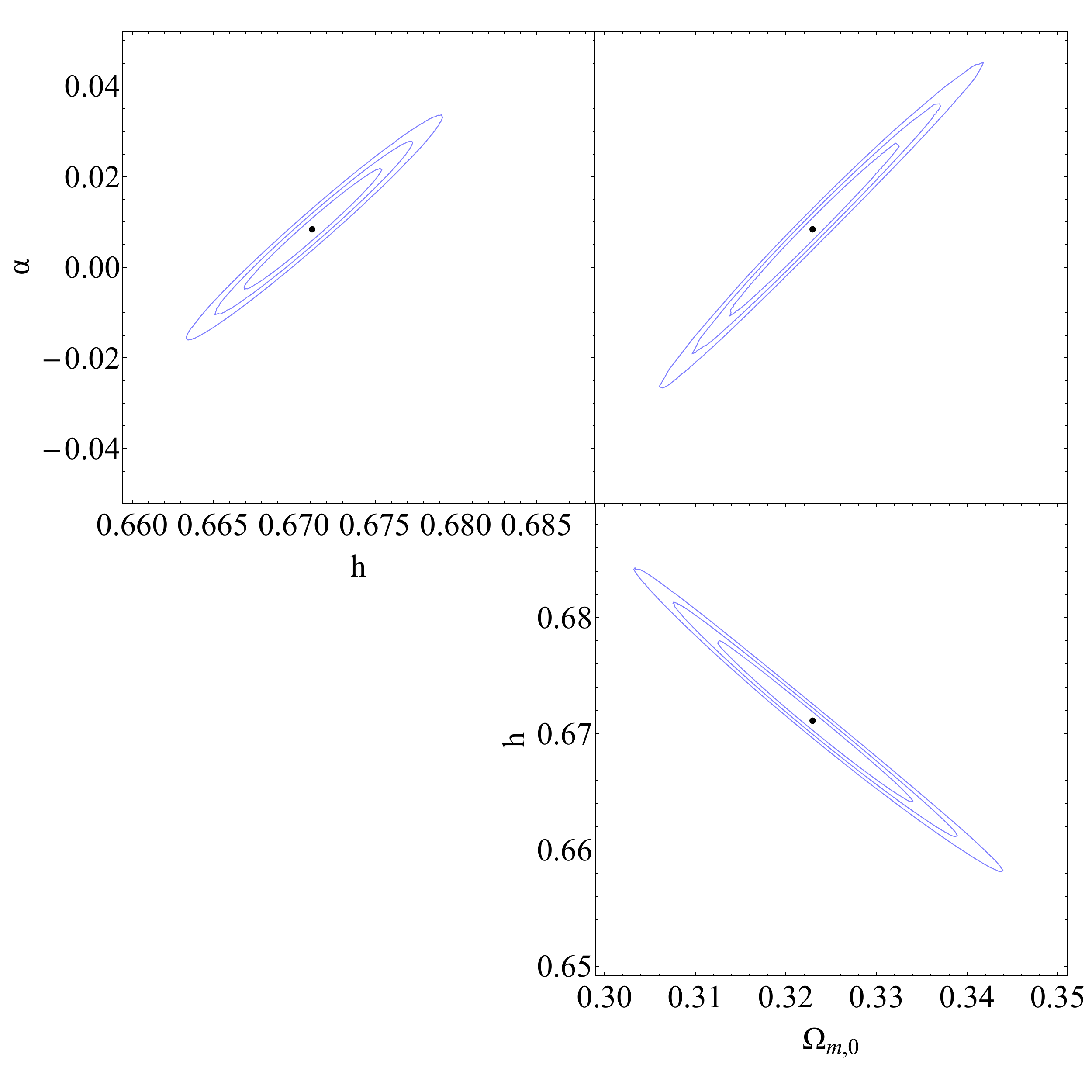}
    \caption{The blue contours in the $h-\alpha$, $\Omega_{m,0}-h$ and $\Omega_{m,0}-h$ diagrams represent the projected $1-3\sigma$ confidence regions based on the CMB+BAO data for the sudden-leap model (sLCDM). The best-fit value for the transition occurring at $a_{s}=0.299\pm0.048$, along with the corresponding best-fit values of $\Omega_{m,0}$, $h$ and $\alpha$, can be found in Table \ref{table2}.}
    \label{Data3}
\end{figure}

 The resulting contours\endnote{Some additional contour plots are presented in Appendix \ref{AppendixΑ} (Figure \ref{DataA2}).} are shown in Figure \ref{Data3}.

 \subsection{Real Data: Pantheon+}\label{subsec:3.3}

Type Ia supernovae, characterized by the absence of a spectral line of hydrogen and the presence of an absorption line attributed to singly ionized silicon, result from the explosion of a white dwarf in a binary system that surpasses the Chandrasekhar limit due to gas accretion from a companion star.

Importantly, type Ia supernovae exhibit a nearly constant absolute luminosity at the peak of their brightness, denoted by an established absolute magnitude of approximately $M \approx -19$. As a result, the distance to a type Ia supernova can be deduced through the observation of its apparent luminosity. By concurrently measuring the apparent magnitude and the light curve, it becomes feasible to predict the corresponding absolute \mbox{magnitude \cite{amendola_tsujikawa_2010}}.

Brighter supernovae exhibit broader light curves (flux or luminosity of the supernova as a function of time). It is important to note that when referring to the universal absolute magnitude of type Ia supernovae hereafter, it is implied that the magnitude has been appropriately adjusted to account for the light curve width.

When considering the luminosity distance $d_{L}$ measured in megaparsecs (Mpc), the concepts of absolute magnitude, apparent magnitude and luminosity distance can be formally related as follows \cite{amendola_tsujikawa_2010}:

\begin{equation}\label{Magn}
\mu \equiv m - M = 5 \log_{10}\left(\frac{d_{L}}{\text{Mpc}}\right) + 25
\end{equation}

Here, $\mu$ represents the distance modulus, which quantifies the difference between the apparent magnitude $m$ and the absolute magnitude $M$ of an object. 

In the context of modified gravity, an abrupt transition in $G_{\text{eff}}$ results in a luminosity distance as predicted by (\ref{eq:solve}). In the event that the luminosity peak is proportional to the absolute magnitude $M$ \cite{Gaztanaga:2001fh} (see Equation (\ref{Magn})), a sudden increase in the effective gravitational constant $G_{\text{eff}}$ would result in a supernova exhibiting diminished brightness relative to the predictions derived from conventional scenarios \cite{Perivolaropoulos:2023iqj}.

The Pantheon+ dataset comprises a collection of $1550$ type Ia supernovae and $1701$ corresponding light curves, spanning a redshift range of $0.001<z<2.26$ \cite{Brout:2022vxf}. To analyze the Pantheon+ data, we adopt the methodology outlined in the study of \mbox{Brout et al. (2022) \cite{Brout:2022vxf}.}

 If we denote the $1701\times 1701$ covariance matrix as $[C_{stat+syst}]$, which is provided with the Pantheon+ data including both statistical and
systematic uncertainties, and if we begin with the minimization a $\chi^2$ distribution

\begin{equation}
    \chi^{2}=\textbf{Q}^{T} [C_{stat+syst}]^{-1}\textbf{Q}
\end{equation}

then due to degeneracy, it becomes impossible to estimate $H_0$ as there is a correlation between $H_0$ and the absolute magnitude of SnIa, denoted as $M$.

The vector $\textbf{Q}$ represents a quantity with 1701 components
and each one is  defined as
\begin{equation}
    Q_{i}=m_{B_{i}}-M-\mu_{model}(z_{i};\Omega_{m,0},h,\alpha,z_{s})
\end{equation}

The predicted distance modulus is denoted as $\mu_{model}$, which is obtained using the assumed sudden-leap model. If the luminosity distance $d_{L}$ is given in  (\ref{eq:solve}), then
 \begin{equation}\mu_{model}(z_{i};\Omega_{m,0},h,\alpha,z_{s})=5\log_{10}\bigg{(}\frac{d_{L}(z_{i};\Omega_{m,0},h,\alpha,z_{s})}{Mpc}\bigg{)}+25\end{equation}

The estimate of $H_0$ was not possible in the first Pantheon sample \cite{Pan-STARRS1:2017jku} because of the degeneracy between $H_0$
and the SnIa absolute magnitude $M$. 

To resolve the degeneracy issue in the Pantheon+ dataset, a modification was made to the vector $\textbf{Q}$ by incorporating the distance moduli of SnIa in Cepheid hosts $\mu_{Ceph}$, which can constrain $M$ independently. Thus, the modified
vector $\textbf{Q}'$ is \cite{Brout:2022vxf}

\begin{equation} 
 Q_{i}' =
  \begin{cases}
    m_{B_{i}}-M-\mu^{Ceph}(z_{i})    & \quad if\quad i\in\, Ceipheid\,\, hosts \\
     
    m_{B_{i}}-M-\mu_{model}(z_{i};\Omega_{m,0},h,\alpha,z_{s}) & otherwise
  \end{cases}
  \end{equation}
 
 \noindent and $\mu^{Ceph}(z_{i})$ is the corrected distance modulus of the Cepheid host of the $i^{th}$ SnIa, which is measured independently in the
context of the SH0ES distance ladder with Cepheid calibrators \cite{Riess:2021jrx}. Thus, the degeneracy between $M$ and $H_0$ is shattered and the parameters $M, H_0, \Omega_{m,0},\alpha,z_{s}$ can be estimated by implementation of the 
  \begin{equation}
  \chi^{2}_{Pant}=\textbf{Q'}^{T} [C_{stat+syst}]^{-1}\textbf{Q'}
  \end{equation}

\begin{figure}[H] 
%    \centering
    \includegraphics[width=0.98\textwidth]{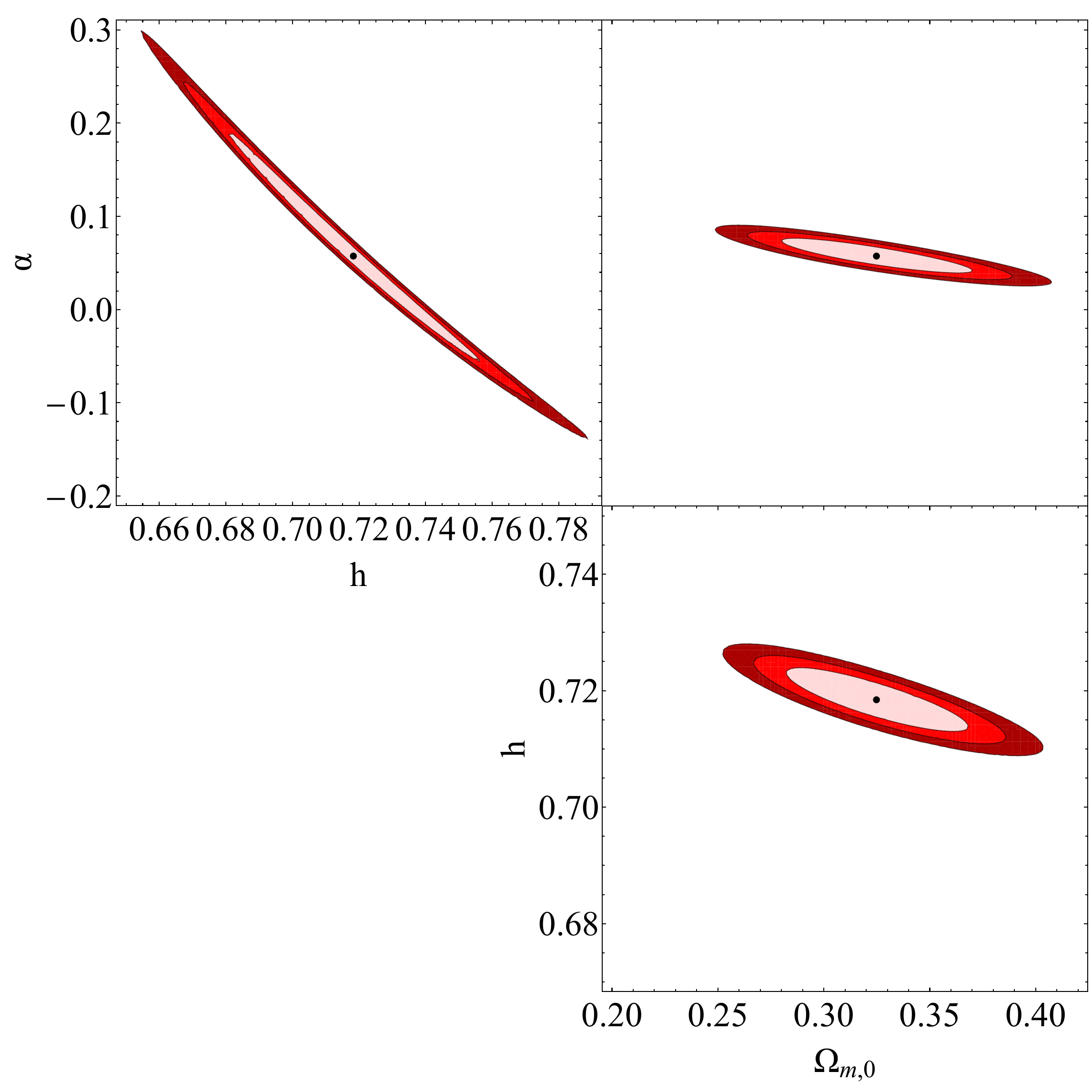}
    \caption{{The red} %MDPI: figure 7 is not after its first citaiton, please correct. %We have corrected the placement of Figure 7!
 contours in the $h-\alpha$, $\Omega_{m,0}-h$ and $\Omega_{m,0}-h$ diagrams represent the projected $1-3\sigma$ confidence regions obtained from analyzing the Pantheon+ data within the sudden-leap model (sLCDM). The best-fit values for the transition occurring at $a_{s}=0.995\pm 0.004$ and $M =19.24\pm 0.03$, along with the corresponding best-fit values of $\Omega_{m,0}$, $h$ and $\alpha$, can be found in Table \ref{table2}. }
    \label{Data4}
\end{figure}

The resulting contours\endnote{Some additional contour plots are presented in Appendix \ref{AppendixΑ} (Figure \ref{DataA3}).}  are shown in Figure \ref{Data4}.

\subsection{Combined Data}\label{subsec:3.4}

\begin{figure}[H] 
%    \centering
    \includegraphics[width=1\textwidth]{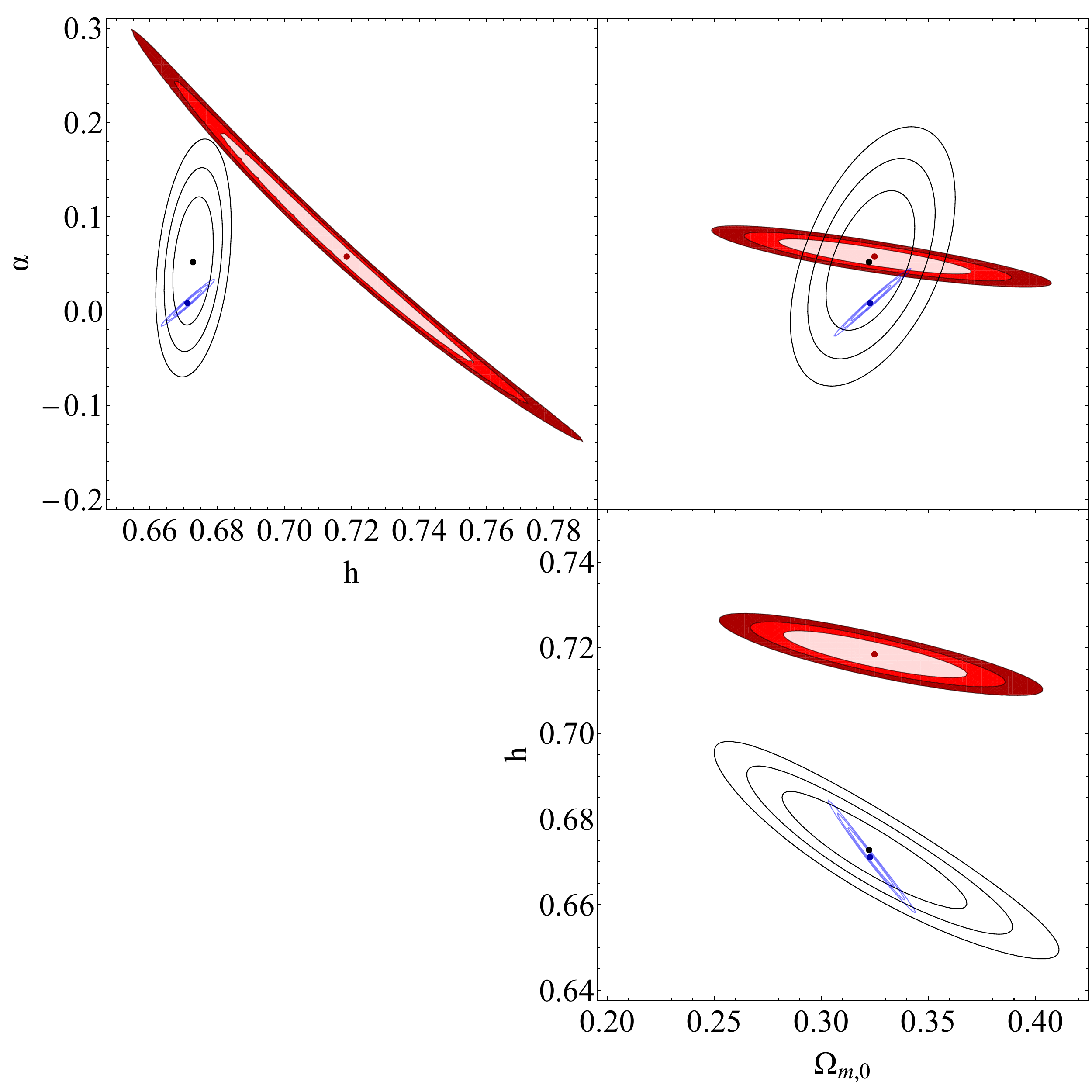}
    \caption{A combined plot including the contours of Figure \ref{Data1} (\textit{black contours}, which correspond to the confidence regions of the mock data of the standard sirens for the sudden-leap model); Figure \ref{Data3} (\textit{blue contours}, which correspond to the confidence regions of the assemblage of $CMB,\,BAO$ data of the sudden-leap model); and {Figure \ref{Data4}} %MDPI: figure 7 citaiton  is after figure 8, figure citation order is wrong, please correct all figure citaiton to numerical order in whole paper.% We have corrected the placement of all figures to numerical order!
 (\textit{red contours}, which correspond to the confidence regions of the Pantheon+ data of the sudden-leap model), in contrast to one another. }
    \label{Data6}
\end{figure}

\begin{figure}[H] 
%    \centering
    \includegraphics[width=1\textwidth]{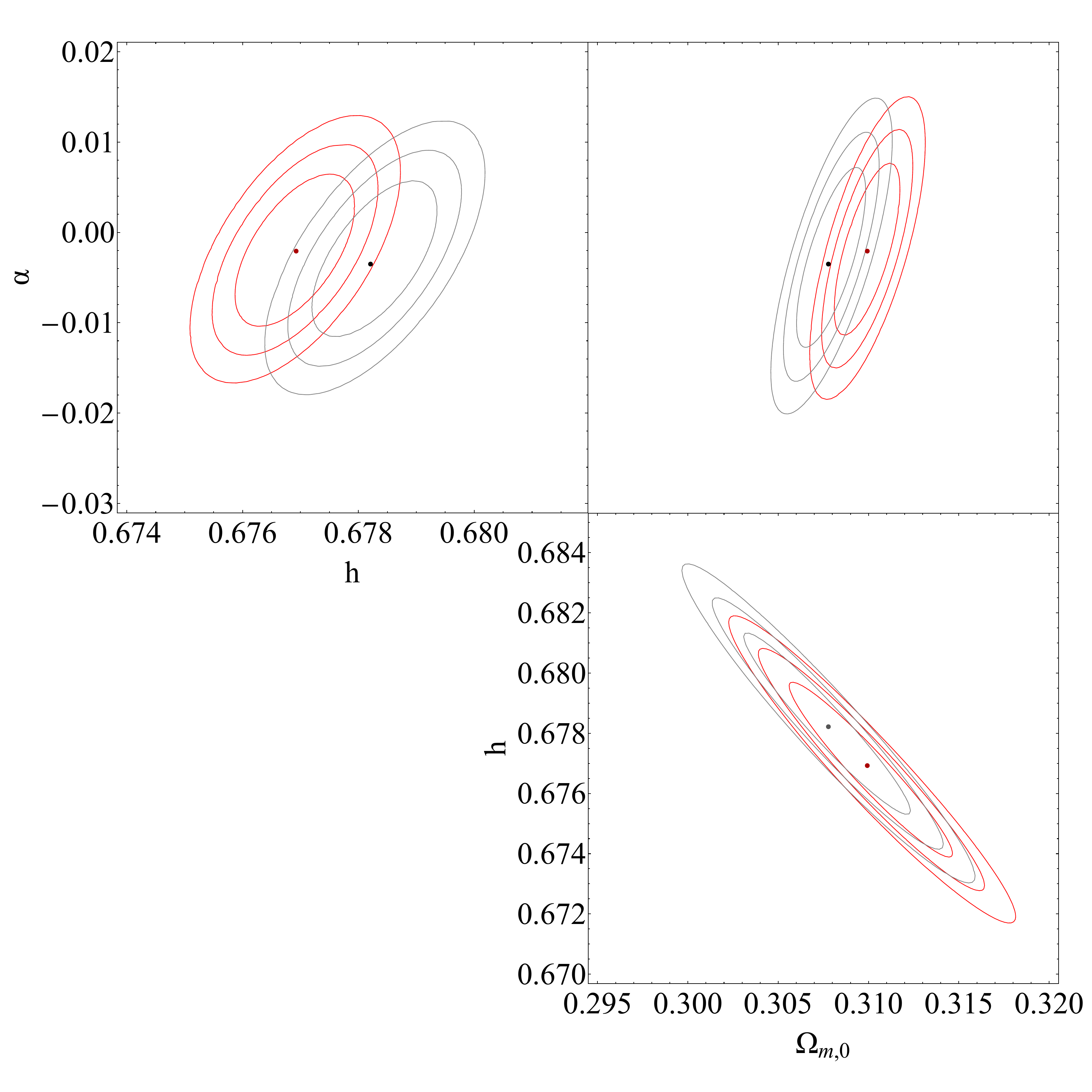}
    \caption{The red contours observed in the $h-\alpha$, $\Omega_{m,0}-\alpha$ and $\Omega_{m,0}-h$ diagrams represent the projected $1-3\sigma$ confidence regions obtained through the analysis of the combined $\chi^{2}$ function within the sudden-leap model (sLCDM). This combined $\chi^{2}$ function comprises contributions from $\chi^{2}_{\text{sirens}}$, $\chi^{2}_{\text{BAO}}$, $\chi^{2}_{\text{Panth}}$ and $\chi^{2}_{\text{CMB}}$. The best-fit values for the transition occurring at $a_{s}=0.812\pm 0.2$ and the absolute magnitude $M=-19.43\pm 0.015$, as well as the corresponding best-fit values for $\Omega_{m,0}$, $h$, and $\alpha$, are provided in Table 2. Additionally, the gray contours reflect the contributions from $\chi^{2}_{\text{BAO}}$, $\chi^{2}_{\text{CMB}}$ and $\chi^{2}_{\text{Panth}}$ for the corresponding best-fit values of $a_{s}=0.789 \pm 0.2$ and $M=-19.43\pm 0.015$. }
    \label{Data11}
\end{figure}

 The previously presented figures are compared in Figure \ref{Data6}. Additionally, by minimizing the $\chi^2$ and considering a range defined as $\chi^2_{\text{min}}\pm\Delta \chi_{n-\sigma}^2$ (as described in Appendix \ref{AppendixB}), we can determine the projected confidence regions for each pair of parameters by fixing the parameters that are not directly studied to their best-fit values. The confidence levels of $1\sigma$, $2\sigma$ and $3\sigma$ are indicated by the red and gray contours in Figure \ref{Data11}.

When combining the CMB, BAO, Standard Sirens and Pantheon+ data, the corresponding $\chi^2$ distribution is obtained by summing the individual $\chi^2$ contributions:

\begin{equation}\label{tot}
\chi^{2}=\chi^{2}_{\text{sirens}}+\chi^{2}_{\text{BAO}}+\chi^{2}_{\text{Panth}}+\chi^{2}_{\text{CMB}}
\end{equation}

Specifically, we set the values of $a_s = 0.812 \pm 0.2$ and $M = -19.43 \pm 0.015$, and additionally fix $\Omega_{m,0} = 0.31 \pm 0.005$ or $h = 0.677 \pm 0.003$ or $\alpha = -0.001 \pm 0.005$ for each diagram, respectively. This  approach is consistent across all three diagrams. For example, in the $h-\alpha$ diagram, the fixed best-fit values are $\Omega_{m,0} = 0.31 \pm 0.005$, $a_s = 0.812 \pm 0.2$ and the absolute magnitude $M = -19.43 \pm 0.015$.

Table \ref{table2} provides the details of the best-fit values. To construct the $h-\alpha$, $\Omega_{m,0}-\alpha$ and $\Omega_{m,0}-h$ diagrams presented in Figure \ref{Data11}, certain parameters are fixed. Figure \ref{Data11} illustrates the contours in \textit{gray}, representing the $\chi^2$ distribution as defined in Equation (\ref{tot1}). These contours correspond to the combined contributions from $\chi^2_{\text{BAO}}$, $\chi^2_{\text{Panth}}$ and $\chi^2_{\text{CMB}}$ given by:

\begin{equation}\label{tot1}
\chi^2 = \chi^2_{\text{BAO}} + \chi^2_{\text{Panth}} + \chi^2_{\text{CMB}}
\end{equation}

The resulting contours\endnote{Some additional contour plots are presented in Appendix \ref{AppendixΑ} (Figure \ref{DataA4}).} are shown in Figure \ref{Data11}.

\subsection{Results}\label{subsec:3.5}

The generated mock data are obtained by assuming the Planck18 values (Planck Collaboration, 2018 \cite{Planck:2018vyg}), which explains the close proximity of the best-fit values for $\Omega_{m,0}$ and $H_{0}$ derived from both standard sirens and the combination of CMB+BAO data \mbox{(Figure \ref{Data6}).}
The case where the value of the transition amplitude is  $\alpha\neq 0$ appears to be disfavored by the data, as the value $\alpha=0$ consistently falls within the $1\sigma$ range in all cases. It is important to highlight that the analysis of the Pantheon+ data suggests a potential quasi-degeneracy between the Hubble constant dimensionless parameter $h$ and the transition amplitude  $\alpha$. Furthermore, the Pantheon+ dataset indicates that the best-fit value for the scale factor is $a_s = 0.995\pm0.004$ ($z_s = 0.005$), and the corresponding luminosity distance at $z_s$ is approximately $d_L(z_s)\approx20.96$ Mpc, as determined by \mbox{Equation (\ref{eq:solve})}. Notably, this luminosity distance value is close to the estimate of a sudden change in the intrinsic luminosity distance of SnIa by \cite{Perivolaropoulos:2023tdt,Perivolaropoulos:2023iqj,Alestas:2021nmi,Perivolaropoulos:2021bds}, which predicts a value of approximately \mbox{$20$ Mpc.}

The present theoretical framework  of the sudden-leap model is characterized by the formation of a true vacuum bubble. This phenomenon arises from a phenomenological first-order phase transition of a scalar field, exemplified by Equation (\ref{equat4}). The transition extends over an approximate luminosity distance of $20.96$ Mpc. Within the confines of this true vacuum bubble, there is a transition in the effective gravitational constant, represented by $G_{\text{eff}}$. Consequently, the encounter of bound systems \cite{Perivolaropoulos:2016nhp} or gravitational waves with the boundary of this vacuum bubble gives rise to profound physical effects akin to those associated with a sudden cosmological singularity.

The assessment of accuracy in parameter estimation entails the evaluation of the ratio between the projected one-dimensional likelihoods and their corresponding best-fit values. This ratio acts as a quantification of the constraining power demonstrated by a particular set of measurements, centered on their respective best-fit values. 

The inclusion of the Pantheon+ data alongside the CMB and BAO data appears to result in a degradation of the predicted accuracy at the best-fit value of $h$ and $\Omega_{m,0}$ for the sudden-leap model. This can be attributed to a plausible quasi-degeneracy between the parameters $h$ and $\alpha$. Conversely, the addition of standard sirens into the parameter-estimation process results in an enhancement of the accuracy for the sudden-leap model in all cases considered (see Table \ref{table3}).

 \begin{table}[H] 
\caption{{Accuracy at the sudden-leap model}.}
\label{table3}
%\newcolumntype{C}{>{\centering\arraybackslash}X}
\setlength{\tabcolsep}{3.6mm}
\begin{tabular}{C{8cm}C{1cm}C{1cm}C{1cm}}
\toprule
{ \textbf{Data:}} %MDPI: is bold necessary?.%Yes, it is.
 & \boldmath{$\frac{\Delta M}{M}$}  & \boldmath{$\frac{\Delta \Omega_{m,0}}{\Omega_{m,0}}$} & \boldmath{$\frac{\Delta h}{h}$}  \\
\midrule
\textbf{ Standard Sirens} & --- &7.4\%& 1.2\%  \\ 

\textbf{CMB +BAO} & --- & 2.8\%  &0.6\% \\ 

\textbf{Pantheon+} &0.2\% &6.2\% &3.2\%\\ 
\textbf{Standard Sirens+Pantheon+} &0.1\% & 4.6\%  &0.6\% \\
\textbf{CMB +BAO+Pantheon+} &0.1\% & 2.9\%  &0.9\% \\

\textbf{Standard Sirens+CMB+BAO+Pantheon+} & 0.1\%&1.6\% &0.4\% \\ 
%Entry 2		& Data			& Data \textsuperscript{1}\\
\bottomrule
\end{tabular}
%\noindent{\footnotesize{\textsuperscript{1} Tables may have a footer.}}
\end{table}
%\textbf{Table 3:}  :

%\begin{adjustbox}{width=12 cm,left}
%\begin{tabular}{cccccc}  \toprule
%  \midrule
% 
% \hline
%
%
%
%
%\bottomrule
%\end{tabular}
%\end{adjustbox}\\

To identify the best model for future observations, we can use the Akaike Information Criterion (AIC). Given a set of models, the AIC helps us select the model that best describes the data. The criterion estimates the expected, relative information loss between the fitted model and the
observed data. The AIC factor is expressed as \cite{burnham2003model}:

\begin{equation}
    AIC=2k-2\ln(\hat{\mathcal{L}})
\end{equation}

The likelihood function could be defined as $\mathcal{L}\approx e^{-\frac{1}{2}\chi^{2}}$ and has a maximum value of $\hat{\mathcal{L}}\approx e^{-\frac{1}{2}\chi_{min}^{2}}$, and the term $k$ represents the number of  parameters in the model (\mbox{see \cite{burnham2003model})}. When comparing a set of models, it is not the absolute value of the AIC that is important, but rather the difference between the AIC values of the corresponding pair of models. In the case of a set of models that includes both the $\Lambda$CDM model and the sudden-leap model, we use the AIC differences, or $\Delta_{i}=AIC_{sLCDM}-AIC_{\Lambda CDM}$, to assess the empirical support for each model \cite{burnham2003model,Shi:2012ma}.

A $\Delta_{i}$ value between 0 and 2 indicates \textit{substantial} empirical support for the $i$-model, while a value between 4 and 7 suggests \textit{considerably less} support. If $\Delta_{i}$ exceeds 10, the empirical support for the $i$-model is \textit{essentially none}. We can compare the hypothesized model with the $\Lambda$CDM model.

\begin{figure}[h] 
%    \centering
    \includegraphics[width=1\textwidth]{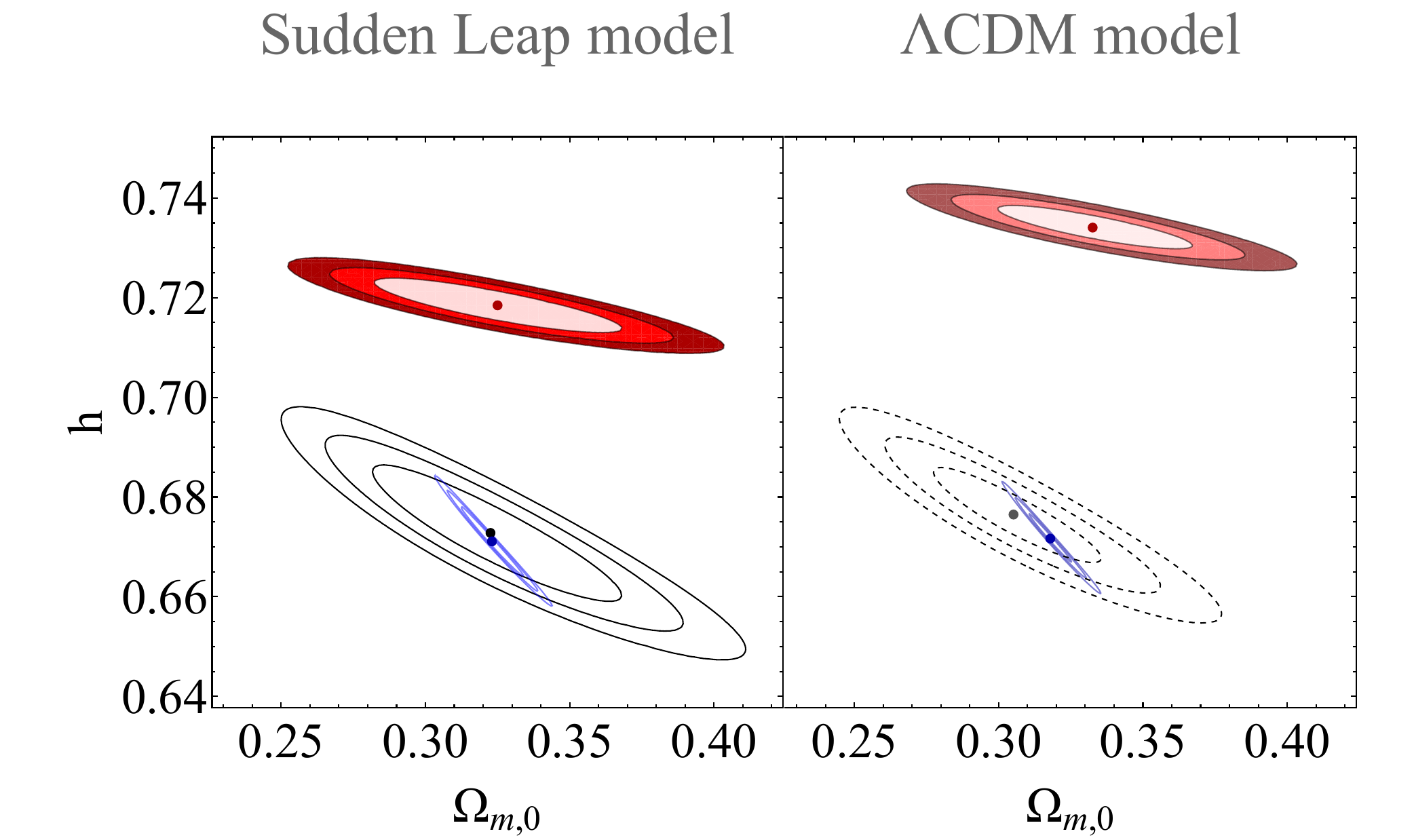}
    \caption{{The resulting} %MDPI: figure 10 missing citaiton, please add, and make sure all figure citaiton are in numerical order.% Confirmed!
 \textit{red} contours, which correspond to the confidence regions of the Pantheon+ data, the \textit{blue} contours to the CMB+BAO data, and the \textit{black} contours to the SS mock data of the sudden-leap model, in contrast to the corresponding confidence regions of the (\textit{red hue}) Pantheon+, the (\textit{blue hue}) CMB+BAO and the (\textit{dashed}) $SS$ mock data of the $\Lambda$CDM model (see Appendix \ref{AppendixC}). }
    \label{Data01}
\end{figure}

To evaluate the relative probability of the $i$-model occurring, we set $e^{-\frac{1}{2}\Delta_{min}}\equiv 1$ for the estimated best model, and use the ratio $e^{-\frac{1}{2}\Delta_{i}}$ \cite{burnham2003model}. \\
\vspace{-9pt}

%\textbf{Table 4:} 

%\begin{adjustbox}{width=12cm,left}
%\begin{tabular}{cccccc}  \toprule
%  \midrule
%
%
%\bottomrule
%\end{tabular}
%\end{adjustbox}\\

\begin{figure}[H] 
%    \centering
    \includegraphics[width=1\textwidth]{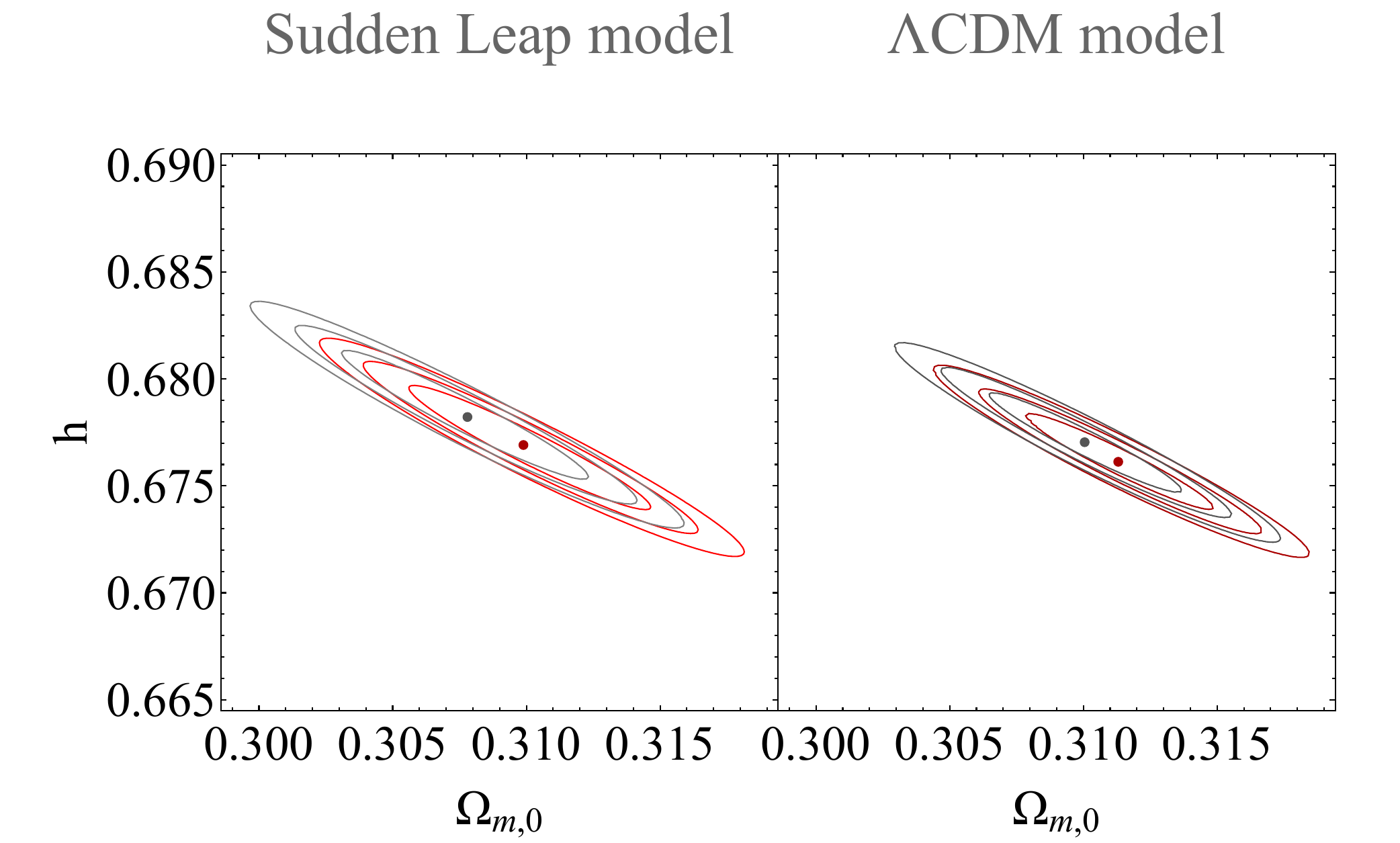}
    \caption{{The resulting} %MDPI: figure 11 missing citaiton, please add, and make sure all figure citaiton are in numerical order.% Confirmed!
 \textit{contours} that correspond to the confidence regions sketched around the minimum of $\chi^{2}_{Pant}+\chi^{2}_{\text{BAO}}+\chi^{2}_{\text{CMB}}$ (\textit{gray}) and $\chi^{2}_{Pant}+\chi^{2}_{\text{BAO}}+\chi^{2}_{\text{CMB}}+\chi^{2}_{\text{sirens}}$ (\textit{red}) of the sudden-leap model in contrast to the corresponding confidence regions sketched around the minimum of $\chi^{2}_{Pant}+\chi^{2}_{\text{BAO}}+\chi^{2}_{\text{CMB}}$ (\textit{gray hue}) and $\chi^{2}_{Pant}+\chi^{2}_{\text{BAO}}+\chi^{2}_{\text{CMB}}+\chi^{2}_{\text{sirens}}$ (\textit{red hue}) of the $\Lambda$CDM model (see Appendix \ref{AppendixC}). }
    \label{Data02}
\end{figure}

 For example, the \textit{sudden-leap model} is about 0.264 times as probable as the $\Lambda$CDM model in minimizing the information loss of Pantheon+ data. The current datasets does not seem to favor the sudden-leap model over the $\Lambda$CDM model in any case (see also Figures \ref{Data01} and \ref{Data02}).

The standard sirens, $\chi^{2}_{\text{sirens}}$, when they are added to the data, with $\chi^{2}_{Pant}+\chi^{2}_{\text{BAO}}+\chi^{2}_{\text{CMB}}$, improve the accuracy of the parameter estimation. This improvement would have been larger if the consistency between the standard sirens and the Pantheon+ data was higher. Notice that for the generation of the  mock siren data we have assumed \plcdm model parameters, which are not fully consistent with the Pantheon+ best-fit \lcdm parameter values. This tension tends to increase the uncertainty of the best-fit parameter values when these datasets are combined.

Even though the increase in constraint level is relatively small (Figure \ref{Data11}), the standard sirens have systematic errors completely independent from the rest of the data \cite{Belgacem_2018}. For these reasons, standard sirens will have a key role in the parameter constraining in \mbox{the future}. \\

\section{{Conclusions and Discussion} %MDPI: please confirm the section title. %We confirm!
}\label{sec:IV}

It has been suggested that the Hubble tension, i.e., the disparity in the Hubble constant measurements at low and high redshifts, could potentially be accounted for by an abrupt transition in the effective gravitational constant taking place within a range of redshifts less than $z_s< 0.01$ \cite{Marra:2021fvf}. Herein, we explored the consequences of such a hypothesis by considering an underlying scalar--tensor theory of gravity and identifying the effects of such a transition on propagating gravitational waves and on cosmological data, including mock standard siren data. We considered a gravitational transition driven by a phenomenological first-order transition of a scalar field $\Phi(z)$, represented as $\Phi(z)=\Phi_{0}[1+\Delta_{1}(\alpha) \Theta(z-z_{s})]$. Such a scalar field in the context of a scalar--tensor theory may lead to a gravitational transition and a Hubble flow of the form $H(z)\propto [1+\sigma \Theta(z-z_s)]$.

This conjecture has the potential to resolve the Hubble tension \cite{Abdalla:2022yfr,Marra:2021fvf,Alestas:2022xxm,Kazantzidis:2020tko} in the context of a weak sudden cosmological singularity involving geodesic completeness. Such singularities have been rigorously explored in diverse studies \cite{Fernandez-Jambrina:2006tkb,Fern_ndez_Jambrina_2021,Perivolaropoulos:2016nhp,Barrow_2004}. They naturally incorporate the abrupt transition in the Hubble flow as described by the functional \mbox{form (\ref{hztrans}).}

A physical mechanism for such a gravitational transition may involve the formation of a true vacuum bubble with a scale of about $20\, Mpc$. In the context of a scalar--tensor theory, this vacuum bubble would include a distinct value of the effective gravitational constant $G_{\text{eff}}$. Consequently, the expansion of the bubble through bound systems or gravitational waves could be realized with cosmological data as a type II sudden cosmological singularity. 

We have shown that the transition and its effects on the Hubble flow would manifest itself as a detectable pattern within gravitational waveforms, akin to the expectations from a sudden cosmological singularity. These patterns emerge due to the additional modified gravity term
denoted as $\alpha_T$, which modifies the friction term of the gravitational wave evolution equation, contributing to gravitational wave amplitude variations, as illustrated in Figures \ref{SFS2} and \ref{SFS3}.

Such a gravitational transition model has been constrained in the context of our analysis using both real cosmological data (Pantheon+, BAO and CMB) and mock simulated standard siren data generated in the context of a \plcdm model. 
In this context, constraints were imposed on the  standard parameters $M$ (SnIa absolute magnitude), $\Omega_{m,0}$ and $h$, as well as on the sudden-leap transition model parameters $\alpha$ (the transition amplitude) and $z_{s}$ (the corresponding redshift). The examination of the correlation among these parameters may shed light on the intricate dynamics and potential links between the scalar field transition, the Hubble flow and the corresponding cosmological parameters.

Future experiments, such as those deploying the ET for detecting standard sirens, hold promise for constraining modified theories of gravity \cite{Belgacem_2018,Belgacem_2019}. In specific variants of modified gravity theories, a discontinuity in the gravitational constant could engender a discrepancy between light propagation and gravitational waves.  With the emergence of third-generation gravitational-wave observatories, for instance, the ET \cite{Zhao_2011,Maggiore:2019uih,Breschi:2022ens,Califano:2022syd} or LISA \cite{Klein:2015hvg,article007,Mangiagli:2022niy,Mangiagli:2022elu}, it will be feasible to directly examine scenarios encompassing gravitational transitions utilizing the luminosity distance $d_{L}^{\text{gw}}$ of gravitational waves \cite{Maggiore:2019uih} and comparing it with the luminosity distance at the same redshift obtained with electromagnetic waves.

We have imposed constraints on the sudden-leap model parameters and demonstrated that it does not appear to fit the considered cosmological data more effectively than the $\Lambda$CDM model (see, e.g., Table \ref{table4}). However, this model has the potential to resolve the Hubble tension via a mechanism that is based on a change in physics that is distinct from other proposed approaches. Thus, this class of models remains worthy of further investigation using both gravitational-wave and electromagnetic-wave cosmological data.

\begin{table}[H] 
\caption{{The $AIC$ differences between the $\Lambda$CDM and the sudden-leap model (sLCDM).}}
\label{table4}
\setlength{\tabcolsep}{2.2mm}
%\newcolumntype{C}{>{\centering\arraybackslash}X}
\begin{tabular}{C{8cm}C{5cm}}
\toprule
{  \textbf{Data}} %MDPI: 	is bold necessary?.%Yes,it is!
 & \boldmath{$\Delta=AIC_{sLCDM}-AIC_{\Lambda CDM}$}    \\
  \midrule

\textbf{ Standard Sirens} & 1.94 \\ 

\textbf{CMB +BAO} & 3.19 \\ 

\textbf{Pantheon+} &2.66 \\ 
\textbf{Standard Sirens+Pantheon+} &3.82 \\ 
\textbf{CMB +BAO+Pantheon+} &3.72 \\

\textbf{Standard Sirens+CMB+BAO+Pantheon+} & 3.9\\ 

\bottomrule
\end{tabular}
%\noindent{\footnotesize{\textsuperscript{1} Tables may have a footer.}}
\end{table}

The potential of the sudden-leap transition class of models to resolve the Hubble tension is demonstrated by the quasi-degeneracy between the parameters $h$ and $\alpha$, which exhibit a significant correlation and are difficult to discriminate using the Pantheon+ dataset (see Figure \ref{Data4}). Thus, the introduction of the transition amplitude parameter $\alpha$ has the potential to change the best-fit value of the Hubble parameter of the Pantheon+ data without a significant change in the quality of fit of the transition model, in contrast with the standard \lcdm model. The identified redshift favored by the Pantheon+ data  for the transition is $z_s=0.005$, where the luminosity distance approximates to $d_L(0.005)\approx 20.96\,Mpc$. This is consistent with previous studies mildly %Please confirm that the intended meaning has been retained.
favoring a sudden shift in the intrinsic SnIa luminosity, around $20\,Mpc$ \cite{Perivolaropoulos:2023iqj}, and a transition in the projected level of anisotropy of the SnIa absolute magnitudes, within approximately $30\, Mpc$ \cite{Perivolaropoulos:2023tdt}. A gravitational transition taking place at these parameter values \cite{Perivolaropoulos:2022khd} could potentially resolve the Hubble tension as well as the observed tension in $S_{8}\equiv \sigma_{8}\sqrt{\frac{\Omega_{m,0}}{0.3}}$ \cite{Marra:2021fvf}.

We have also demonstrated that the integration of standard siren data into the sudden-leap model results in an increase in the precision of the measurements for the parameters $\Omega_{m,0}$ and $h$, thereby diminishing their uncertainties from $(2.9\%,0.9\%)$ to $(1.6\%, 0.4\%)$. The significant merit of incorporating standard siren data is attributed to the fact that the systematic errors related to them are entirely different from the electromagnetic wave data.

There are two main directions for the extension of the present analysis:
\begin{itemize}
\item 
Construction of a more detailed theoretical model that can induce the gravitational transition discussed in the present analysis. Such a model could be based on a scalar--tensor theory involving a false vacuum decay transition or a transition in time due to features of the potential determination of the dynamics of the scalar field.%Please confirm that the intended meaning has been retained.
\item 
Consideration of additional astrophysical and/or cosmological data at redshifts \mbox{$z<0.01$} (distances less than $40Mpc$) to impose constraints on such a gravitational transition or identify hints for its existence (see, e.g., \cite{Alestas:2021nmi}).
\item 
The investigation of the effects of other types of singularities on the propagation of gravitational waves and on the luminosity distance as measured by either electromagnetic or gravitational waves. For instance, in the context of a generalized sudden cosmological singularity the $r=3$ derivative of the scale factor diverges. In this case, we have $\frac{\ddot{a}(t)}{a(t)}$$\sim$$[1+\alpha\Theta(t-t_{s})]$, while the scale factor $a$ and its derivative $\dot{a}$ are continuous. Alternatively, a type III singularity can be constructed by admitting that $a(t)$$\sim$$[1+\alpha \Theta(t-t_{s})]$, or even a $w$-singularity by allowing that \mbox{$w(t)=w[1+\alpha \delta(t-t_s)]$.}
\item 
The consideration of alternative gravitational-wave mock data corresponding to other future gravitational-wave observatories including LISA \cite{LISA:2017pwj,Caprini:2015zlo} and the comparison of the constraints that can be imposed on the parameters of the sudden-leap transition model with those of the ET considered in the present analysis. LISA is projected to detect frequencies spanning from $0.1\,mHz$ to $10^{-1}\,Hz$, a markedly distinct frequency range from that of the ET. Within this spectrum of frequencies, the anticipation is to detect a myriad of GW sources \cite{article004}, including Galactic binaries \cite{Breivik:2017jip,article005,article006,Lau:2019wzw}, binary systems hosting stellar-origin black holes \cite{Sesana:2016ljz}, extreme-mass-ratio inspirals (EMRIs) \cite{Babak:2017tow}, mergers of massive black hole binaries (MBHBs) even at high redshifts \mbox{(up to $z$$\sim$$10$) \cite{Mangiagli:2022niy}} and possibly stochastic GW backgrounds \cite{Caprini:2018mtu}.
\end{itemize}

\

\authorcontributions{E.A.P. made contributions to the writing, methodology, data analysis, and literature investigation. L.P. contributed to the conceptualization, writing, and general supervision of the project. All authors have thoroughly reviewed and approved the published version of the manuscript. %MDPI: please add this part.
}

%\authorcontributions{For research articles with several authors, a short paragraph specifying their individual contributions must be provided. The following statements should be used ``Conceptualization, X.X. and Y.Y.; methodology, X.X.; software, X.X.; validation, X.X., Y.Y. and Z.Z.; formal analysis, X.X.; investigation, X.X.; resources, X.X.; data curation, X.X.; writing---original draft preparation, X.X.; writing---review and editing, X.X.; visualization, X.X.; supervision, X.X.; project administration, X.X.; funding acquisition, Y.Y. All authors have read and agreed to the published version of the manuscript.'', please turn to the  \href{http://img.mdpi.org/data/contributor-role-instruction.pdf}{CRediT taxonomy} for the term explanation. Authorship must be limited to those who have contributed substantially to the work~reported.}
\funding{This article is based upon work from COST Action CA21136 - Addressing observational tensions
in cosmology with systematics and fundamental physics (CosmoVerse), supported by COST (European Cooperation in Science and Technology).
This project was also supported by the Hellenic Foundation for Research and Innovation (H.F.R.I.), under the ”First call for H.F.R.I. Research Projects to support Faculty members and Researchers and the
procurement of high-cost research equipment Grant”
(Project Number: 789). %MDPI: please add this part.
}

\conflictsofinterest{The authors declare no conflict of interest.%MDPI: please add this part.
} 
\appendixtitles{yes} % Leave argument "no" if all appendix headings stay EMPTY (then no dot is printed after "Appendix A"). If~the appendix sections contain a heading then change the argument to~"yes".

\appendixstart

\appendix

\section[\appendixname~\thesection]{Sudden-Leap Model (sLCDM)}
\label{AppendixΑ}
In this Appendix we show some additional contour plots for parameters not shown in the main text.
 \begin{figure}[H] 
%    \centering
    \includegraphics[width=0.75\textwidth]{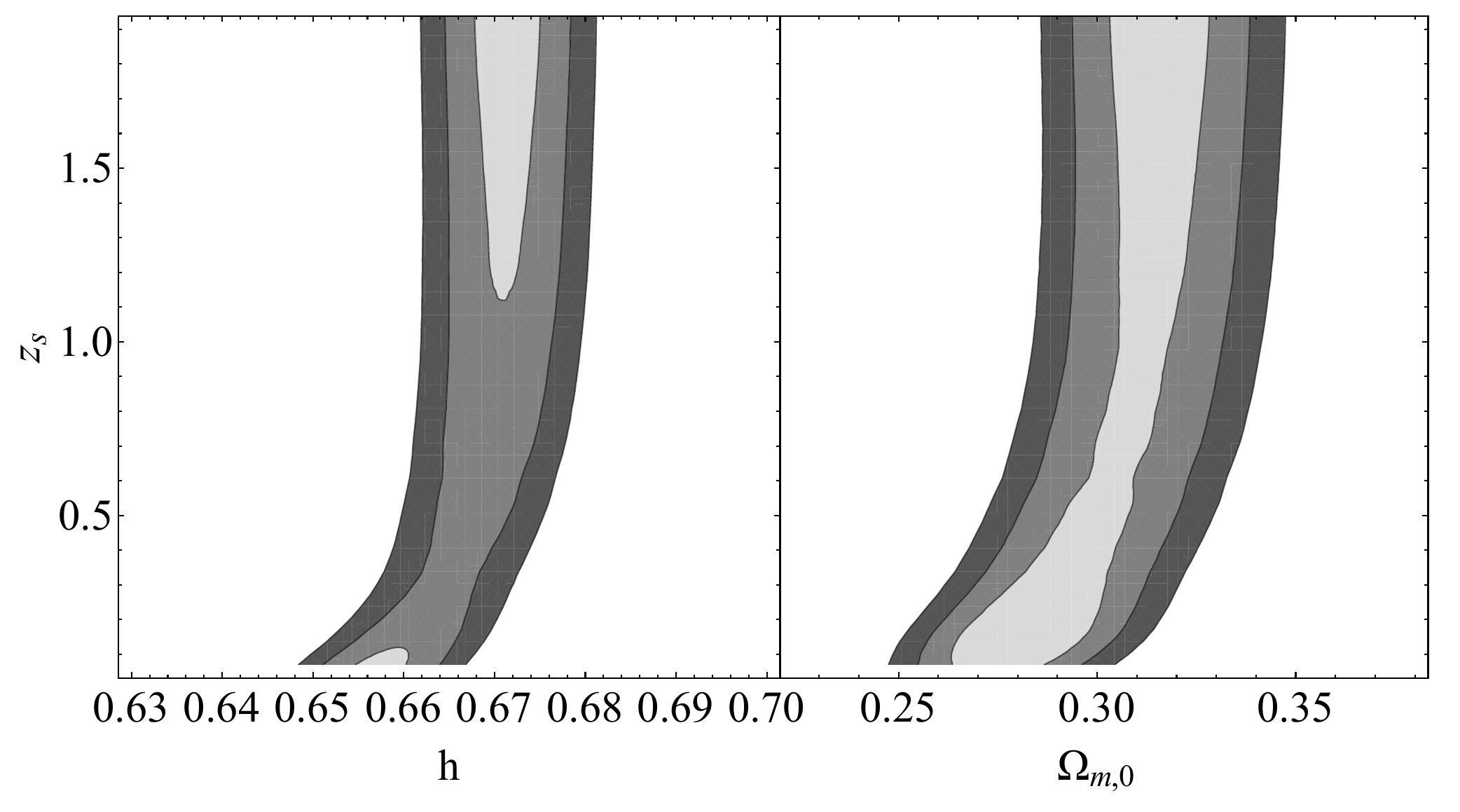}
    \caption{\textit{Gray contours} represent the $1-3\sigma$ confidence regions of the standard siren data  for the corresponding best-fit values (Table \ref{table2}). }
    \label{DataA1}
\end{figure}
 
\vspace{-9pt}

\begin{figure}[H] 
%    \centering
    \includegraphics[width=0.85\textwidth]{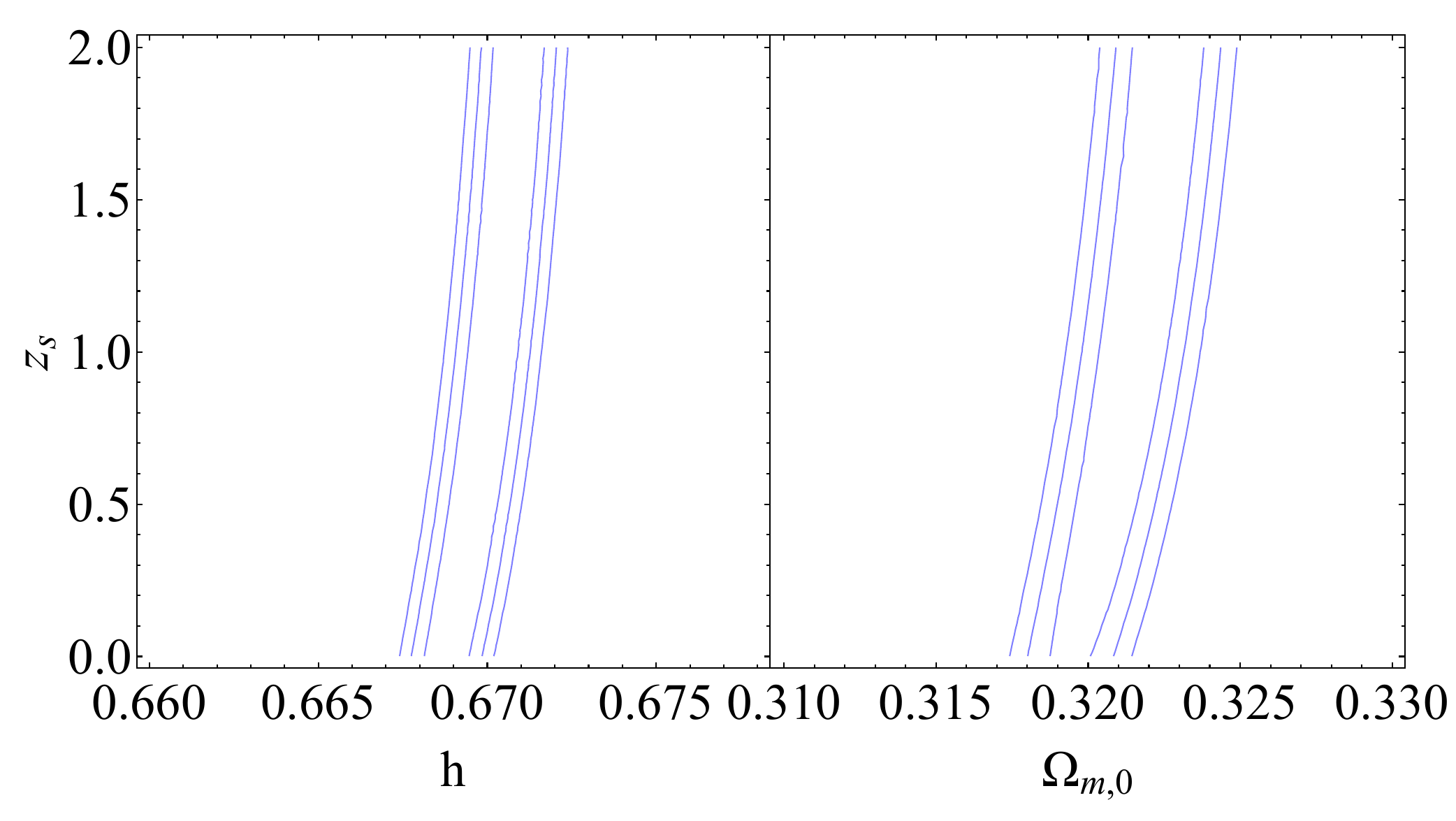}
    \caption{\textit{Blue contours} represent the projected  $1-3\sigma$ confidence regions in $h-z_{s},\,\Omega_{m,0}-z_s$ diagrams of the CMB+BAO data for the sudden-leap model (sLCDM); see Table \ref{table2}.}
    \label{DataA2}
\end{figure}

\vspace{-6pt}

 \begin{figure}[H] 
%    \centering
    \includegraphics[width=0.7\textwidth]{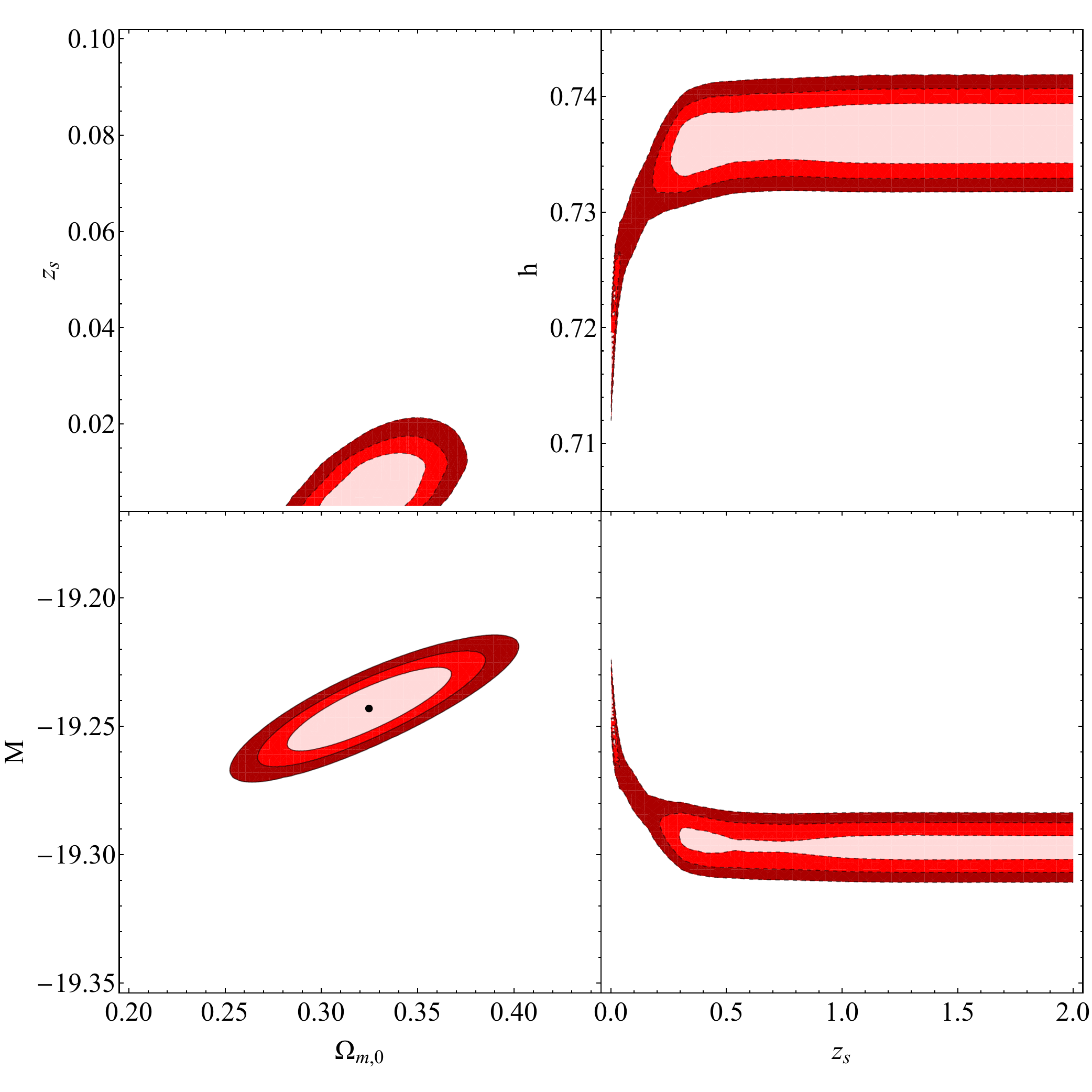}
    \caption{\textit{Red contours} represent the $1-3\sigma$ confidence regions of the $\Omega_{m,0}-z_{s},\,\Omega_{m,0}-M,\,z_{s}-M,\,z_{s}-h$ for the Pantheon+ data and their corresponding best-fit values.}
    \label{DataA3}
\end{figure}
\vspace{-9pt}
\begin{figure}[H] 
%    \centering
    \includegraphics[width=0.8\textwidth]{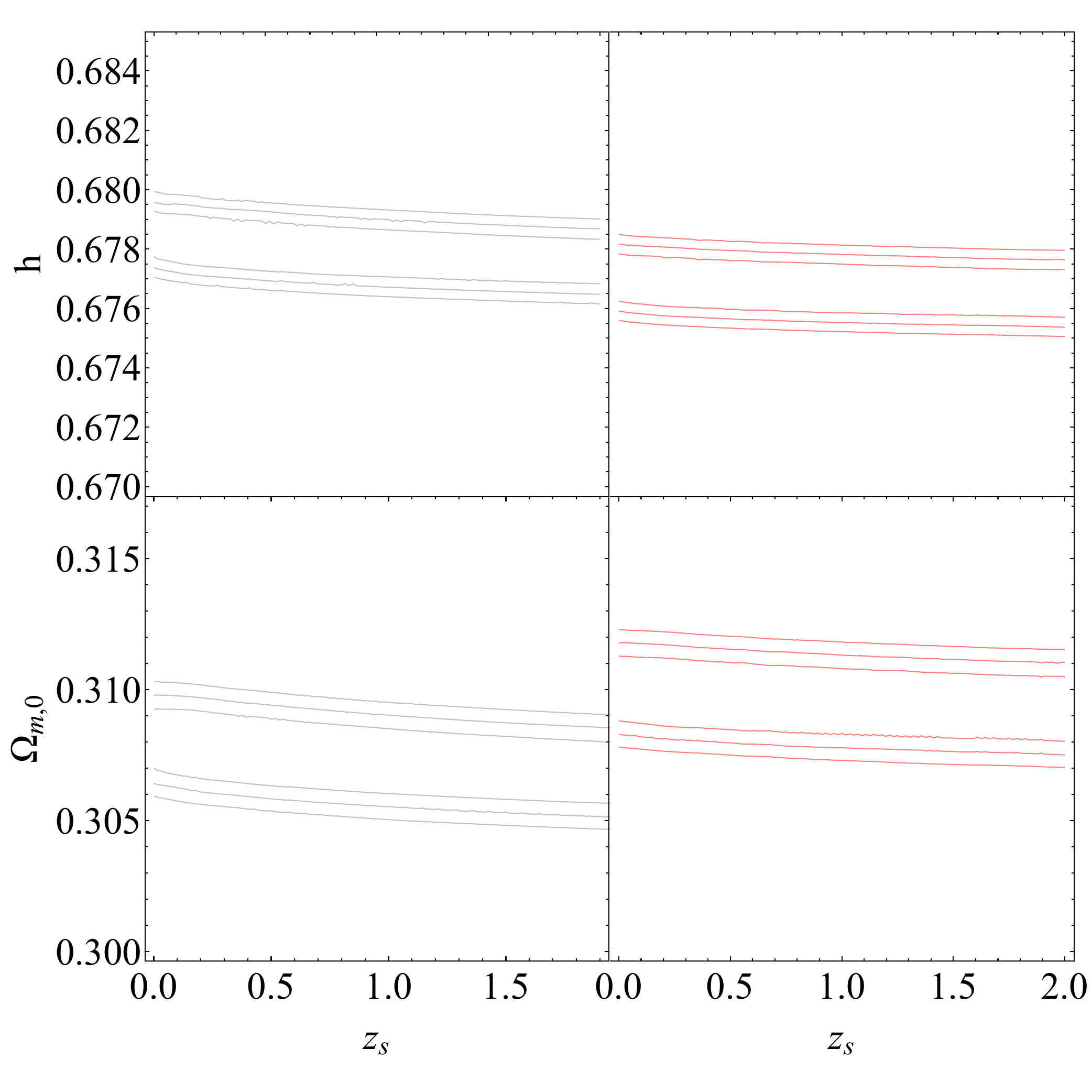}
    \caption{\textit{Red contours} represent the projected $1-3\sigma$ confidence regions in $z_{s}-h,\, z_{s}-\Omega_{m,0}$ diagrams of the $\chi^{2}=\chi^{2}_{\text{sirens}}+\chi^{2}_{\text{BAO}}+\chi^{2}_{\text{Panth}}+\chi^{2}_{\text{CMB}}$ distribution, while the \textit{gray} contours represent the  $\chi^{2}=\chi^{2}_{\text{BAO}}+\chi^{2}_{\text{Panth}}+\chi^{2}_{\text{CMB}}$. }
    \label{DataA4}
\end{figure}

%\newpage

\section[\appendixname~\thesection]{Maximum Likelihood Method}
\label{AppendixB}

A brief overview of the maximum likelihood method is presented. To establish the confidence regions for an unknown parameter in a given distribution, the utilization of the $\chi^{2}$ distribution is required. Let $\{x_{i}\}$ denote a collection of measurements, $\textbf{x}_{obs}$ represent the associated data vector and $\textbf{x}=\textbf{x}_{theory}-\textbf{x}_{obs}$ be the difference between the theoretical and observed data vectors. The $\chi^{2}$ distribution is defined \cite{dodelson2020modern,article411,Theodoropoulos:2021hkk,Verde_2010}

\begin{equation}
     \chi^{2}(\theta)=\textbf{x}^{T}[C]^{-1}\textbf{x}=[x_{i,th}(\theta)-x_{i,obs}]\big{(}[C]^{-1}\big{)}_{ij}[x_{j,th}(\theta)-x_{j,obs}]
 \end{equation}
where $[C]_{ij}\equiv C(x_{i},x_{j})$ is the covariance and is defined as:
\begin{equation}
    [C] = 
 \begin{pmatrix}
  \sigma^{2}_{1} & \rho_{12} \sigma_{1}\sigma_{2} & \cdots& \rho_{1n} \sigma_{1}\sigma_{n}\\
   \rho_{21} \sigma_{1}\sigma_{2}&\sigma^{2}_{2}  & \cdots& \rho_{2n} \sigma_{2}\sigma_{n}\\
  \vdots  & \vdots  & \ddots & \vdots  \\
    \rho_{n1} \sigma_{n}\sigma_{1}& \rho_{n2} \sigma_{n}\sigma_{2} & \cdots& \sigma^{2}_{n} 
 \end{pmatrix}
\end{equation}

The likelihood function $\mathcal{L}$ of an $n$-dimensional random variable $\{X_{i}\}$ is defined as $\mathcal{L}\big{(} \{x_{i}\}|\theta\big{)}=f(\{x_{i}\};\theta)$, where $f$ denotes  the corresponding probability density function of $\{X_{i}\}$ and the unknown parameter $\theta$. In the present context, the likelihood function is 

\begin{equation}
    \mathcal{L}\big{(} \{x_{i}\}|\theta\big{)}= N e^{-\frac{1}{2}\chi^{2}(\theta)}
\end{equation}
 where $N$ denotes a normalization constant. The likelihood function is used to estimate the most likely ``true'' parameters $\theta$, which are those that sit at a higher value of the likelihood function. In this context, the minimal value of the $\chi^{2}$ distribution, represented as $\chi^{2}_{min} \equiv \chi^{2}(\hat{\theta})$, aligns with the maximum of the likelihood function.  The maximum likelihood estimator $\hat{\theta}$ may exist or not and is not necessarily unique.

By defining the curvature matrix of the likelihood function, the \textit{Fisher matrix}  is a measure of how rapidly $\chi^{2}$ varies away from its minimum in the corresponding directions. If the corresponding components of the Fisher matrix are large in some directions, then the likelihood changes rapidly and the data are constraining, i.e., the resulting uncertainties in the parameters will be small enough. The Fisher matrix is defined:

  \begin{equation}
      \mathcal{F}_{ij}\equiv-\bigg{<}\frac{\partial^{2}\ln\mathcal{L}}{\partial\theta_{i}\partial\theta_{j}}\bigg{>}
  \end{equation}
  
The Cramer--Rao inequality posits that $\sigma_{ij}\geq \sqrt{([\mathcal{F}]^{-1})_{ij}} $, and when all parameters are estimated simultaneously, $\sigma_{\theta_{i}}\geq\sqrt{([\mathcal{F}]^{-1})_{ii}} $. In the case that the likelihood is Gaussian, we have $\sigma_{ij}= \sqrt{([\mathcal{F}]^{-1})_{ij}}$ and $\sigma_{\theta_{i}}=\sqrt{([\mathcal{F}]^{-1})_{ii}} $ \cite{article411,Theodoropoulos:2021hkk,Verde_2010}.

When dealing with $M$ parameters in the Fisher matrix and requiring 2D plots, marginalization is applied over the uninteresting parameters ($M-2$ parameters). This involves inverting $[\mathcal{F}]$ and selecting the rows and columns corresponding to the desired parameters, then inverting the resulting ``projected'' submatrix \cite{amendola_tsujikawa_2010,Verde_2010}.

When the Fisher matrix is singular, no inverse exists, precluding marginalization. Singularity of the Fisher matrix implies a linear combination of two or more parameters in the likelihood. For the degenerate parameters $\theta_{1},\theta_{2}$, we postulate a new parameter $\Theta=a\theta_{1}+b\theta_{2}$ to avoid singularity. Notably, a particular case of interest arises when there is a high degree of degeneracy between the two corresponding parameters, referred to as quasi-degeneracy. In such situations, a subsequent redefinition of the parameters could once again be necessitated.

The confidence regions live in a parameter space and by projecting them in 2D, their mappings are confidence ``ellipses''. We use the confidence regions that correspond to 68.27\%, 95.45\% and 99.73\% of the total probability distribution. For example, a \textit{confidence region} of 95.45\% for $\theta$ means that, if we repeat  many times the collection of the data and if for each sample we calculate the confidence region of 95.45\% for $\theta$, by computing 

  \begin{equation}
      \int^{\theta_{+}}_{\theta_{-}}d\theta \,\frac{\mathcal{L}\big{(} \{x_{i}\}|\theta\big{)} P(\theta)}{P(\{x_{i}\})}=0.9545
  \end{equation}
  where $\theta_{-},\theta_{+}$ correspond to the values on either side of each maximum $\hat{\theta}$ and have the same probability, then 95.45\% of the confidence regions will contain the \textit{true value} of $\theta$. The confidence regions are those that ``move'' as they depend on the corresponding collection of data that are chosen each time. 
  
  The one-to-one correspondence between $\mathcal{L}$ and  $\chi^{2}$ helps us to estimate the best-fit values for the parameters of interest and also to define the confidence regions around the best-fit values of the corresponding parameters. A fixed $\Delta \chi^{2}$ corresponds to a certain confidence region in parameter space $M$.
  
  The corresponding $\Delta \chi^{2}$ for a  parameter space $M$ of 3--5 dimensions can be seen {below:} %MDPI: please confirm the format of below array.%We confirm!

%\begin{table}[H] 
%\caption{This is a table caption. Tables should be placed in the main text near to the first time they are~cited.\label{tab1}}
%%\newcolumntype{C}{>{\centering\arraybackslash}X}
%\begin{tabularx}{\textwidth}{CCC}
%\toprule
%\textbf{Title 1}	& \textbf{Title 2}	& \textbf{Title 3}\\
%\midrule
%Entry 1		& Data			& Data\\
%Entry 2		& Data			& Data \textsuperscript{1}\\
%\bottomrule
%\end{tabularx}
%\noindent{\footnotesize{\textsuperscript{1} Tables may have a footer.}}
%\end{table}
%  
%    \begin{adjustbox}{width=\linewidth,left}

\setlength{\tabcolsep}{7.25mm}
\noindent \begin{tabular}{L{2cm}L{2cm}L{2cm}L{2cm}}
\toprule
 \multicolumn{4}{c}{$\Delta \chi^{2}_{n-\sigma}$} \\
\midrule
$dim(M)$ & 68.27\% & 95.45\% & 99.73\%\\
\midrule
3   & 3.52674   &8.02488&   14.1564\\
 4&  4.71947  & 9.71563  &16.2513\\
5&5.8876 & 11.3139&  18.2053\\

 \bottomrule
\end{tabular}
%  \end{adjustbox}\\
  
  \vspace{6pt}
  
   \textls[-25]{The projected  confidence regions are generated by the intervals of the form \mbox{$\chi^{2}_{min}\pm\Delta \chi^{2}_{n-\sigma}$.}} In the direction of constraining the data, we demonstrate a schematic overview of the procedure \cite{dodelson2020modern}:

     $$ Covariance\, Matrix \to Fisher \, Forecast\to Parameter\, Constraints$$

\section[\appendixname~\thesection]{ $\Lambda$CDM Model}
\label{AppendixC}

Here, we present a concise summary of the constraints and the corresponding accuracy obtained through the data analysis performed for the $\Lambda$CDM model.

\begin{table}[H] 
\caption{{$\chi^{2}$ {distributions} %MDPI: two table conbined, please confirm the change.% We confirm!
 and the best-fit parameters can be seen below.}}
\label{tableB1}
\footnotesize
%\newcolumntype{C}{>{\centering\arraybackslash}X}
\setlength{\tabcolsep}{2.48mm}

\begin{adjustwidth}{-\extralength}{0cm}
%\centering %% If there is a figure in wide page, please release command \centering
\begin{tabular}{C{1.5cm}C{1.5cm}C{1.5cm}C{3cm}C{4cm}C{4cm}}
\toprule
 { \textbf{Sirens}} %MDPI: is bold necessary?.% Yes,please!
 & \textbf{CMB +BAO} & \textbf{Pantheon+}& \textbf{Standrad Sirens+Pantheon+}& \textbf{CMB+BAO+Pantheon+} & \textbf{Sirens+CMB+BAO+Pantheon+} \\
\midrule
1015.07  &6.39 &1522.98& 2575.73  & 1570.97 & 2586.54\\ 
%Entry 2		& Data			& Data \textsuperscript{1}\\
\midrule
 \multicolumn{3}{c}{\textbf{Data:}} & \boldmath{$M$}  & \boldmath{$\Omega_{m,0}$} & \boldmath{$h$}  \\
 
 \multicolumn{3}{c}{\textbf{Standard Sirens}} & --- & 0.305  $\pm$0.019& 0.676 $\pm$ 0.006 \\ 

 \multicolumn{3}{c}{\textbf{CMB +BAO}} & --- & 0.318 $\pm$ 0.006  &0.672 $\pm$ 0.004 \\ 

 \multicolumn{3}{c}{\textbf{Pantheon+}}&$-$19.25 $\pm$ 0.03 &0.333 $\pm$ 0.018 &0.734 $\pm$ 0.01\\ 

 \multicolumn{3}{c}{\textbf{Standard Sirens+Pantheon+}} & $-$19.41 $\pm$ 0.01 &0.298 $\pm$ 0.012 &0.683 $\pm$ 0.004\\ 

 \multicolumn{3}{c}{\textbf{CMB +BAO+Pantheon+}} & $-$19.43 $\pm$ 0.01 &0.31  $\pm$ 0.005 &0.677 $\pm$ 0.004 \\
 \multicolumn{3}{c}{\textbf{CMB+BAO+Pantheon++Standard Sirens}} & $-$19.43 $\pm$ 0.01 &0.311 $\pm$ 0.004 &0.676 $\pm$ 0.003 \\ 
\bottomrule
\end{tabular}
\end{adjustwidth}
%\noindent{\footnotesize{\textsuperscript{1} Tables may have a footer.}}
\end{table}

%\textbf{Table C1:} }

%\begin{adjustbox}{width=\linewidth,left}
%\begin{tabular}{cccccc}  \toprule
%  \midrule
% 
% 
%\hline

%
%\bottomrule
%\end{tabular}
%\end{adjustbox}
%
%
%\begin{adjustbox}{width=\linewidth,left}
%\begin{tabular}{cccccc}  \toprule
%  \midrule
%
% \hline
%
% 
%
%\bottomrule
%\end{tabular}
%\end{adjustbox}\\

\vspace{-9pt}

\begin{table}[H] 
\caption{{Accuracy at the $\Lambda$CDM model (see also {\cite{Belgacem_2018}.})} %MDPI: \hl{} %Please make sure that permission has been obtained and there is no copyright issue. %There are no copyright issues with the table, as it has been generated by us. Its citation is intended to facilitate a comparison with the corresponding table in the reference!
}
\label{tableB2}
\setlength{\tabcolsep}{4.85mm}
%\newcolumntype{C}{>{\centering\arraybackslash}X}
\begin{tabular}{C{7cm}C{1cm}C{1cm}C{1cm}}
\toprule
  {\textbf{Data:}} %MDPI: is bold necessary?.%Yes, please!
 & \boldmath{$\frac{\Delta M}{M}$}  & \boldmath{$\frac{\Delta \Omega_{m,0}}{\Omega_{m,0}}$} & \boldmath{$\frac{\Delta h}{h}$}  \\
 \midrule

\textbf{ Standard Sirens} & --- &6.2\%& 0.9\%  \\ 

\textbf{CMB +BAO} & --- & 1.9\%  &0.6\% \\ 

\textbf{Pantheon+} &0.2\% &5.4\% &1.4\%\\ 
\textbf{Standard Sirens+Pantheon+} &0.1\% &4.0\% &0.6\%\\ 
\textbf{CMB +BAO+Pantheon+} &0.1\% & 1.6\%  &0.6\% \\

\textbf{Standard Sirens+CMB+BAO+Pantheon+} & 0.1\%&1.6\% &0.4\% \\ 
\bottomrule
\end{tabular}
%\noindent{\footnotesize{\textsuperscript{1} Tables may have a footer.}}
\end{table}

%\textbf{Table C2:} :
%
%
%
%\begin{adjustbox}{width=12cm,left}
%\begin{tabular}{cccccc}  \toprule
%  \midrule

%\bottomrule
%\end{tabular}
%\end{adjustbox}

\begin{adjustwidth}{-\extralength}{0cm}

\printendnotes[custom]

%%%%%%%%%%%%%%%%%%%%%%%%%%%%%%%%%%%%%%%%%%
%\end{paracol}

% Please provide either the correct journal abbreviation (e.g. according to the “List of Title Word Abbreviations” http://www.issn.org/services/online-services/access-to-the-ltwa/) or the full name of the journal.
% Citations and References in Supplementary files are permitted provided that they also appear in the reference list here. 

%=====================================
% References, variant A: external bibliography
%=====================================

%\raggedleft
%\bibliography{Bibliography}

%\centering %% If there is a figure in wide page, please release command \centering
\reftitle{References}

%\externalbibliography{yes}
%\bibliography{bibliography.bib}

\PublishersNote{}
\end{adjustwidth}

%%%%%%%%%%%%%%%%%%%%%%%%%%%%%%%%%%%%%%%%%%
%% optional

%% for journal Sci
%\reviewreports{\\
%Reviewer 1 comments and authors’ response\\
%Reviewer 2 comments and authors’ response\\
%Reviewer 3 comments and authors’ response
%}

%%%%%%%%%%%%%%%%%%%%%%%%%%%%%%%%%%%%%%%%%%
\end{document}